\documentclass[9pt,twocolumn,twoside]{osajnl}

\journal{josab} 

\setboolean{shortarticle}{false}

\usepackage{supertabular}
\usepackage[titletoc]{appendix}
\usepackage{array}
\usepackage{siunitx}
\usepackage{cuted}
\usepackage{booktabs}

\newcommand{\be}{\begin{equation}}
\newcommand{\ee}{\end{equation}}
\newcommand{\bs}{\boldsymbol}

\newcolumntype{L}[1]{>{\arraybackslash}p{#1}}
\title{Atomic transitions for adaptive optics}

\author[1,2]{Rui Yang}
\author[2]{Joschua Hellemeier}
\author[2,3,*]{Paul Hickson}

\affil[1]{School of Physics and Astronomy, Yunnan University, South Section, East Outer Ring Road, Chenggong District, Kunming, 650500, China}
\affil[2]{Department of Physics and Astronomy, University of British Columbia, 6224 Agriculture Road, Vancouver BC, V6T 1Z1,Canada}
\affil[3]{National Astronomical Observatories, Chinese Academy of Sciences, A20 Datun Road, Chaoyang District, Beijing, China}

\affil[*]{Corresponding author: hickson@physics.ubc.ca}

\begin{abstract}
Adaptive optics systems using sodium laser guide stars are widely employed at major astronomical observatories. It is natural to ask whether other atomic species might offer advantages. In this paper we review all abundant atoms and ions in the upper atmosphere, including Na, Fe, Mg$^+$, Si$^+$, Ca$^+$, K and also the non-metallic species N, N$^+$, O, H, considering their potential for adaptive optics. Return fluxes for all transitions that can be excited using either one or two wavelengths were computed.  We also considered multi-wavelength emission, comparing the performance of different transitions for polychromatic laser guide star (PLGS) adaptive optics. We find that of all the mesospheric metals, Na is the most suitable for both monochromatic and polychromatic laser guide stars, \textcolor{black}{providing about six times more return flux than the best transitions in Fe}. For high-altitude observatories, excitation at 330 nm in Na should give the highest PLGS performance. Atomic O, N and N$^+$ have strong transitions and very high abundances in the mesosphere. This makes them potential candidates for the generation of intense laser guide stars by amplified spontaneous emission, if a suitable excitation process can be demonstrated.  Direct excitation by CW lasers is impractical as all transitions from the ground state are beyond the atmospheric cutoff. Nevertheless, it may be possible using high-power pulsed lasers. 
\end{abstract}

\setboolean{displaycopyright}{true}

\begin{document}

\maketitle

\section{Introduction}
Adaptive optics (AO) systems are employed by many ground-based telescopes to compensate atmospheric turbulence, thereby greatly improving the resolution of astronomical images \cite{Beckers1993}. Many of these systems employ laser guide star (LGS) beacons, which allow sensing of  atmospheric turbulence in any desired direction, increasing sky coverage \cite{Foy1985}.  Large telescopes typically employ resonant LGS systems in which  transitions in mesospheric atom are excited, typically the $D_2$ transition of neutral atomic sodium.  Although other chemical species have been considered \cite{Happer1994,Telle1996}, sodium is preferred due to its relatively-high abundance and strong resonance lines.

Despite the impressive success of AO systems employing sodium LGS, there are several reasons to reconsider other atoms and ions. The performance of such AO systems is limited by the return flux of the LGS, due to photon shot noise. Advances have been made both in increasing laser power, and improving the efficiency of sodium LGS \cite{Fan2016,Holzlohner2010,Milonni1999,Rampy2015,Bustos2020}. For example, simulations indicate that repumping atoms from the $D_{2b}$ ground state, in combination with circular polarization, can increase the return flux by a factor of $3-4$ \cite{Holzlohner2010}. Nevertheless, any technique that might further increase the return flux would be valuable. 

Several new techniques that might potentially achieve high-intensity LGS are discussed in \cite{Hickson2020} . One possible approach is  amplified spontaneous emission (ASE) \cite{Bustos2018}. In this idea, laser pumping of an abundant atmospheric atomic or molecular species could produce a population inversion that would amplify radiation generated by spontaneous emission. A critical requirement for ASE is a suitable transition for which the optical depth $\tau$, defined as the integral of the absorption coefficient along the propagation path, is negative and has an absolute value that exceeds unity. This is a consequence of the equation of radiative transfer, from which it follows that the gain in intensity is proportional to $\exp(-\tau)$.

\textcolor{black}{Another motivation for considering alternatives to sodium is the problem of atmospheric tilt indeterminacy \cite{Rigaut1992}. Backscattered photons follow the same path as the transmitted beam, so the apparent position of the LGS is independent of wavefront tilt induced by the atmosphere. A number of techniques have been proposed to break this degeneracy (see, for example,  \cite{Esposito1998,Ragazzoni2000} and references therein). A promising approach is that of polychromatic laser guide stars (PLGS) \cite{Foy1995,Pique2006}. In this technique, a laser is used to generate emission at two or more widely-separated wavelengths. Because of differential refraction, the backscattered light propagates along different paths through the atmosphere. Thus one need not rely on natural guide stars to sense the atmospheric tilt. The differential positions of the PLGS at different wavelengths is small and must be measured with high precision. Creating sufficiently-intense polychromatic emission is a challenge.}

\textcolor{black}{There is also a related problem of focus indeterminacy, as the sodium centroid altitude varies on short timescales \cite{Pfrommer2010}. This makes it difficult to separate atmospheric focus variations from other focus errors. This problem primarily affects very-large telescopes as the wavefront error due to focus variations increases in proportion to the square of the telescope aperture diameter  \cite{Davis2006}. }

The present paper examines atoms and ions in the upper atmosphere that might be useful for AO. We consider the abundance and distribution of all relevant species and compute the relative return flux for transitions that can be excited by ground-based lasers. Our aim is to identify potential transitions for monochromatic and PLGS, and to assess the feasibility of ASE. 

\section{Mesospheric  atoms and parameters of interest}
\label{sec:mesosphere}

Meteoric ablation is the primary source of metallic species in the mesosphere and lower thermosphere  (MLT) region of the atmosphere, at altitudes between 80 and 120 km. In this altitude range, the atmospheric density decreases from $10 ^{-5}$ to $10^{-8}$ kg m$^{-3}$ and collision rates are in the range $10^3 - 10^5$ s$^{-1}$. This is also the coldest region of the atmosphere. At mid-latitude sites, the temperature at the mesopause ranges from $178 - 192$ K, at an altitude of $85 - 102$ km, depending on the season \cite{She2000}. 

\textcolor{black}{Typical values of the column density $N$, altitude of the centroid of the density distribution, and the RMS width (vertical extent) for each species are listed in Table \ref{tab:Atomic Parameters}. For sodium, we use recent data by Gardner \cite{Gardner2004} to update earlier values listed by Happer \cite{Happer1994}. The centroid altitude and RMS width for non-metallic species were computed from the MSIS-90 atmospheric model \cite{Hedin1991}. These were evaluated at midnight, January 1 for latitude $+30^\circ$ and longitude $0^\circ$.} 

The major meteoric species are Na, Fe, Mg and Si. There are also two less-abundant species, Ca and K. These metals exist as layers of atoms between about 80 and 105 km and occur primarily as atomic ions at higher altitudes. Below 85 km they form compounds: oxides, hydroxides, and carbonates, which polymerize into nanometer-sized meteoric smoke particles. Above 85 km, photolysis of N$_2$, O$_2$ and H$_2$O by extreme ultraviolet (EUV) radiation leads to high concentrations of nitrogen, oxygen and hydrogen atoms. These atoms attack metallic compounds such as hydroxides and oxides, reducing them and thereby returning the metals to their atomic phase.

\begin{table}[htbp]
\begin{center}
\caption{\bf Parameters of mesospheric atoms and ions} \vspace{4pt}
\begin{tabular}{lSSSr}
\hline
Species & {~~~~~$N$} & {Altitude} &  {RMS width} & Sources \\
& {~~~~~~~m$^{-2}$} & {km} &{km}  \\
\hline\\[-9pt]
Na  & 4.0$\times 10^{13}$ & 91.5 & 4.6 & \cite{Gardner2004} \\
Fe & 10.2\hspace{-1.5mm}$\times10^{13}$ & 88.3 & 4.5 & \cite{Gardner2004} \\
Mg & 1$\times10^{13}$ & 89.4 & 4.2 & \cite{Plane2010,Whalley2011} \\
Mg$^+$ & 8$\times10^{13}$ & 94.6 & 7.0 & \cite{Plane2010,Whalley2011} \\
Si$^+$ & 4$\times10^{13}$ & 111.0 & 10.1 & \cite{Plane2016} \\
Ca & 3.4$\times10^{11}$ & 90.5 & 3.5 & \cite{Gardner2004} \\
Ca$^+$ & 7.2$\times10^{11}$ & 95.0 & 3.6 &\cite{Gardner2004}  \\
K & 4.5$\times10^{11}$ & 91.0 & 4.7 & \cite{Gardner2004} \\
N & 1$\times10^{18}$ & 223.0 & 60.2 & \cite{Aikin2000,Hedin1991} \\
N$^+$ & 1$\times10^{18}$ & {-} & {-} & \cite{Aikin2000} \\
O & 6.5$\times10^{21}$ & 110.3 & 28.7 & \cite{Mlynczak2013,Hedin1991} \\
H & 2.3$\times10^{18}$ & 95.6 & 38.9 & \cite{Mlynczak2014,Hedin1991} \\
\hline
\end{tabular}
\label{tab:Atomic Parameters}
\end{center}
\end{table}

In a typical AO application, a laser is used to excite atoms from the ground state to an excited state. The atom then returns to the ground state via spontaneous and stimulated emission. Only spontaneous emission contributes to backscattered radiation since stimulated emission propagates in the same direction as the laser radiation. The LGS return flux can be computed from the equation of radiative transfer. For an optically-thin medium (optical depth $\tau \ll 1$) the specific intensity $I_\nu$ (W m$^{-2}$ Hz$^{-1}$ sr$^{-1}$) of the backscattered radiation is given by  \cite{Hickson2020}
\be
  I_\nu(\nu) = \frac{h\nu}{4\pi} N A_{21} x \varphi(\nu) = \frac{2h\nu}{\lambda^2} N \sigma_{21}(\nu)  x. \label{eq:Inu}
\ee
Here $\lambda$ is the wavelength, $\nu = c/\lambda$ is the frequency, $c$ is the speed of light in vacuum and $N$ is the column density of atoms of this species. $A_{21}$ is the Einstein A coefficient for the transition and the function $\varphi(\nu)$ is the mean line profile, normalized so that its integral over frequency $\nu$ is unity. The dimensionless parameter $x$ is the occupation fraction of the upper state, i.e. the equilibrium fraction of atoms found in this state under continuous excitation by a laser having specific energy density $\rho_\nu$, and $\sigma_{21}(\nu)$ is the cross section for stimulated emission. The second equality \textcolor{black}{follows} from the relation between stimulated and spontaneous emission, $B_{21}$ = $\lambda^3 A_{21}/8\pi h$, and the definition
\be
  \sigma_{21}(\nu) \equiv \frac{h\nu}{c} B_{21}\varphi_\nu(\nu) = \frac{\lambda^2}{8\pi}A_{21}\varphi(\nu).\label{eq:sigma21}
\ee

Here the subscript 1 denotes the lower (ground) state, 2 denotes the upper state,  and $B_{21}$ is the Einstein B coefficient for stimulated emission. 

For a CW laser having beam area $\mathcal{A}$ in the MLT, the return flux $\Phi$, in units of photons s$^{-1}$ m$^{-2}$ is related to the intensity by
\be
  \Phi = \frac{\mathcal{A}}{z^2} \int_0^\infty \frac{I_\nu(\nu)}{h\nu} d\nu, \label{eq:F}
\ee
where $z$ is the line-of-sight distance to the LGS. 

In the mesosphere, Doppler broadening is the dominant line-broadening process. As the temperature varies within the MLT, the line profile is more precisely written as
\be
  \varphi(\nu)  = \frac{1}{N}\int_0^\infty n(z) \varphi(\nu,z) dz,
\ee
where $n(z)$ is the number density of the atomic species under consideration at distance $z$ along the line of sight. In practice, the function $\varphi(\nu)$ is well-represented by a Gaussian
\be
\varphi(\nu) = \frac{2}{w}\sqrt{\frac{\ln 2}{\pi}} e^{-4\ln 2(\nu-\nu_0)^2/w^2}, \label{eq:line_profile}
\ee
where $w$ is the line width at half maximum intensity. It depends on the characteristic temperature $T$ in the MLT and the mass $m$ of the atomic species, 
\begin{equation}
w = \frac{1}{\lambda_0}\left(\frac{8k_BT\ln2}{m}\right)^{1/2}
\label{eq:Doppler width}
\end{equation}
where $k_B$ is Boltzmann's constant. Adopting $T = 180$ K results in a width for the sodium D line of 1.02 GHz. 

The absorption cross section $\sigma_{12}(\nu)$ is given by
\begin{equation}
\sigma_{12}(\nu) = \frac{g_2}{g_1}\sigma_{21}(\nu), \label{eq:sigma12}
\end{equation}
where $g_1$ and $g_2$ are the statistical weights of the lower and upper state, respectively. Tabulated values of cross sections $\sigma_{21} \equiv \sigma_{21}(\nu_0)$ are computed at the line center $\nu_0$.

Substituting Eqs. (\ref{eq:Inu}) and (\ref{eq:sigma21}) into Eq. (\ref{eq:F}), we obtain
\begin{align}
  \Phi & = \frac{\mathcal{A}}{4\pi z^2}  N A_{21}  \int_0^\infty x(\nu) \varphi(\nu) d\nu  \nonumber \\
  & \simeq \frac{\mathcal{A}}{4\pi z^2}  N A_{21}  x(\nu_0) \varphi(\nu_0) \Delta\nu, \label{eq:F3}
\end{align}
where $\Delta\nu$ is the laser line width. The second equality follows because the line width for a CW laser, and therefore the frequency range over which $x$ is nonzero, is typically much smaller than the Doppler width. From this we see that the atomic quantity that best characterizes the LGS return flux,  is 
\be
  \varepsilon = N A_{21}  x(\nu_0).
\ee
We shall call $\varepsilon$ the \emph{return flux coefficient} for the transition. It has units of photons s$^{-1}$ m$^{-2}$.

Also of interest is the dimensionless product $N\sigma_{21}$, which determines the maximum possible optical depth. Specifically, $\tau$ is bounded by $-N\sigma_{21} \le \tau \le N\sigma_{12}$ \cite{Hickson2020}. Therefore, ASE is not possible if $N\sigma_{21} < 1$.

Several other physical processes also play a role in the response of atoms to laser excitation. Atomic collisions effectively reset atoms that have been excited, either returning them to the ground state, or causing a transition to a different excited state. The latter process makes the atom unavailable for excitation by the laser. At the same time, collisions impart a velocity to the atom, generally pushing it out of the velocity interval that can be excited by the laser (the ``velocity class''). In equilibrium this is balanced by  collisions that move atoms into the velocity class. Those atoms are in the ground state as the collision energies, typically $\sim 0.02$ eV, are not sufficient to excite the atom from the ground state. Therefore any individual atom is effectively ``reset'' to the ground state on a time scale equal to the mean time between collisions. 

Collision rates vary by species and altitude. Rates for the species of interest are calculated in Appendix A, for a range of altitudes. 

In order to be useful for AO, a transition should have a high spontaneous transition rate to the lower state (Einstein $A$ coefficient). Also, as most atoms will be in the ground state, there should be a path to the upper state from the ground state, with a high transition probability (Einstein $B$ coefficient). As the Einstein coefficients are related, it is sufficient to consider transitions that have a large $A$ coefficient.

A second consideration is that the upper state should not have many transitions to metastable states, which have low $A$ coefficients. In such a situation, atoms can become trapped in those states, reducing the number of atoms available for the desired transition. 
Ultimately, atoms trapped in metastable states will be reset at the collision rate. 

The performance of PLGS will critically depend on the wavelengths and return flux of the available transitions. For amplitude $\theta$ of a Zernike mode (except for piston), the ratio of the difference in amplitude $\Delta \theta$ at different wavelengths, is related to the ratio of refracting indexes by \cite{Foy1995}
\begin{equation}
  p \equiv \frac{\theta}{\Delta \theta} = \frac{n_{\lambda_\text{obs}}-1}{n_{\lambda_{1}}-n_{\lambda_{2}}}, 
\end{equation}
where $n_{\lambda_{1}}$ and $n_{\lambda_{2}}$ are the refractive indexes at wavelengths of the measurement and $n_{\lambda_{obs}}$ is the refractive index of the wavelength at which the amplitude of the Zernike mode is determined. The dimensionless ratio $p$ is called the \emph{penalty factor}. Smaller values of $p$ correspond to greater PLGS performance.  We take $\lambda_\text{obs} = 500$ nm as the reference wavelength. 

The impact of return flux and penalty factor can be captured in a single merit function. The signal-to-noise ratio for the measurement of the atmospheric wavefront tilt is proportional to the square root of the return flux and inversely proportional to the penalty factor \cite{Foy1995}. Therefore, $\varepsilon^{1/2}/p$ is a measure of relative PLGS performance. To make this dimensionless, we divide by the same quantity evaluated for a reference system. We thus define the PLGS merit function
\be
  q \equiv \frac{p_0}{p}\left(\frac{\varepsilon}{\varepsilon_0}\right)^{1/2}, \label{eq:q}
\ee
where $\varepsilon_0$ and $p_0$ are reference values. For these we use the 330.298/2208.370 nm transitions in Na I, which is the highest-performance system employing single-photon excitation. In all cases, the return flux coefficient used is that of the weaker line. 

When discussing atomic transitions we use spectroscopic notation and refer to Fe as Fe I, Fe$^+$ as Fe II, etc. We refer to transitions having wavelengths greater than 300 nm as ``visible'' transitions. Shorter-wavelengths are greatly attenuated by the atmosphere so we refer to those as ``vacuum-ultraviolet'' transitions. We refer to transitions having an Einstein $A$ coefficient greater than $10^6$ s$^{-1}$ as ``strong transitions'', and ``weak transitions'' otherwise. This distinction is somewhat arbitrary but serves to generally separate permitted lines from forbidden lines. An excitation scheme that uses two different transitions to reach an excited state will be called a ``two-step'' process. One that uses two photons to excite a single transition, using a virtual intermediate state, will be called a ``two-photon'' process. 

Transition wavelengths described in the text and listed in the tables in the following sections are in air for wavelengths greater than 185 nm and in vacuum otherwise. 

\section{Analysis and Results}
\label{sec:analysis}

The data set employed for this analysis is the atomic database of the National Institute of Standards and Technology (NIST) \cite{Kramida2020}. Complete data for all transitions were downloaded and a computer program was written to apply various criteria to identify transitions of interest for AO. We examined all excited states that could be reached from the ground state by at most two successive excitations using wavelengths greater than 300 nm. For species in which no such transitions are possible, two-photon excitation was considered. For every such state, all possible downward transitions were examined, and the excitation fraction $x$ of the excited state was computed according to the methods presented in Appendix B. A specific energy density of $\rho_\nu = 10^{-12}$ J m$^{-1}$ Hz$^{-1}$ was assumed for the laser excitation. This is at the high end of range of energy densities that can be achieved with current AO lasers. A complete listing of all such transitions that have $A \ge 10^7$ s$^{-1}$ is \textcolor{black}{available from the authors upon request}. From this and the data of Table 1, the return flux coefficient was computed. For PLGS, the penalty factor $p$ and merit function $q$ were computed for all downward transitions. 

Our results are shown in Tables 2 -- 9. Table 2 lists all transitions, for all species, that have a return flux coefficient $\varepsilon > 10^{18}$ photons s$^{-1}$ m$^{-2}$. In this, and subsequent tables, $\lambda_\text{ex}$ refers to the excitation wavelength(s) employed to reach the excited state, and $\lambda$ is the wavelength of the downward transition. Unless otherwise specified, values of $p$ are computed with respect to the smallest wavelength.

We now discuss individual atomic species in more detail. 

\subsection{Sodium}
\begin{figure}[tbp]
\centering
\includegraphics[width=8.4cm]{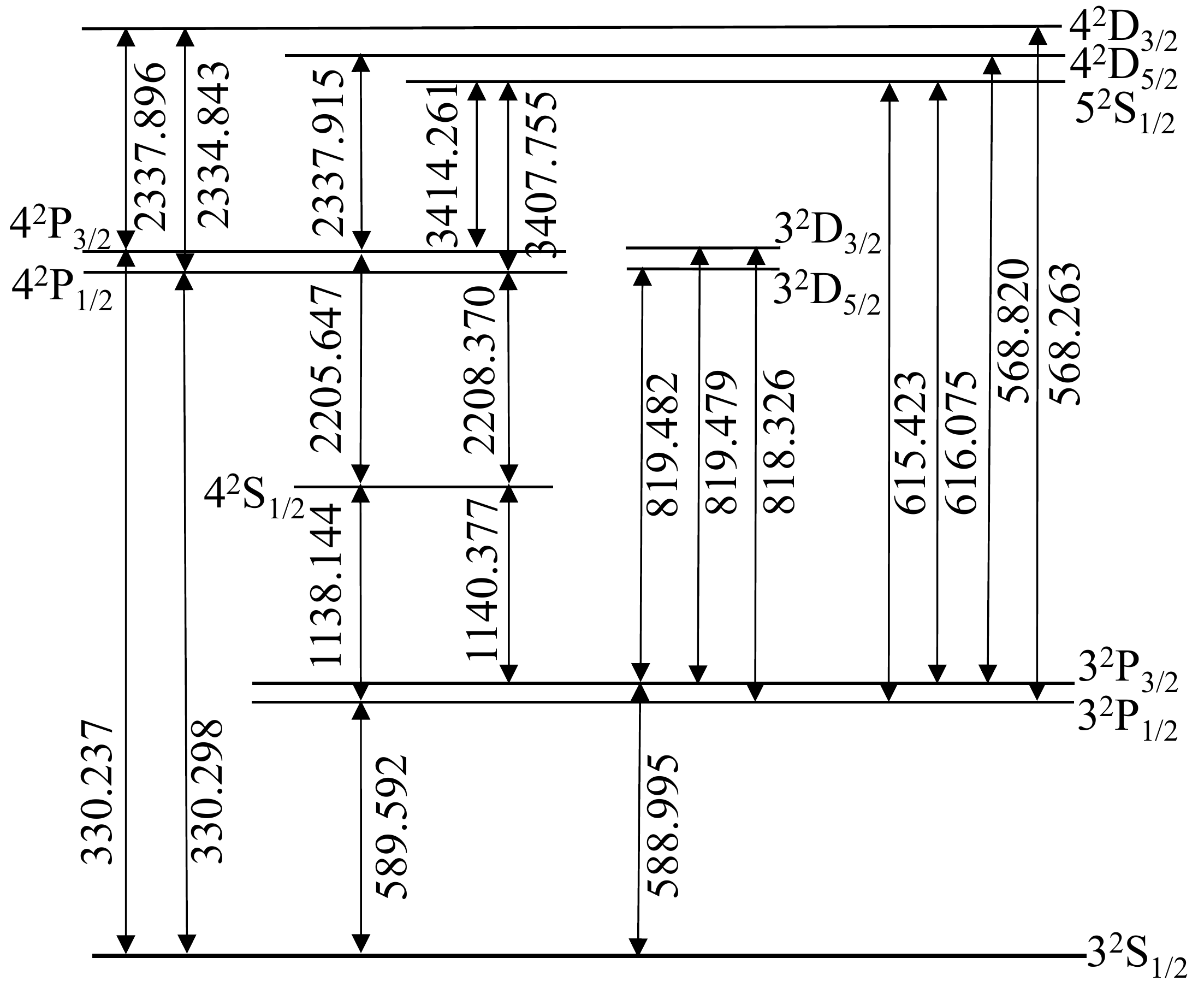}
\caption{Low-lying energy levels of the sodium atom and transitions of interest. }
\label{fig: fig01Na}
\end{figure}

Sodium is most-often employed for the generation of LGS. Neutral sodium is used as it is more abundant than ionized forms.  Averaged over all seasons and latitudes, the observed ratio of Na$^+$ to Na is between 0.2 and 0.25 \cite{Marsh2013}. An energy level diagram of neutral sodium, indicating the lifetimes and wavelengths for selected transitions is shown in Fig. \ref{fig: fig01Na}. For simplicity, fine-structure splitting of the levels is not shown. 

At visible wavelengths, Na I has four strong transitions from the ground state, at 330.237, 330.298, 588.995 and 589.592 nm. The latter two form the sodium D doublet. A fifth transition, at 388.390 nm, has a very-small rate ($A = 6.95\times10^{-4}$ s$^{-1}$). 

The transition between the 3S$_{1/2}$ ground state and the 3P$_{3/2}$ excited state, the D$_2$ line, has been of interest for adaptive optics since 1985 \cite{Foy1985}. Laser-induced excitation followed by spontaneous emission produces the backscattered photons of the LGS. The LGS return flux is limited by the low abundance and small absorption cross section, $N\sigma_{12} \simeq 3$\%. In practice, the D$_{2b}$ line, which involves only transitions between the $F = 2$ lower and $F = 1, 2$ and 3 upper sub-levels, is used for AO. A CW laser can excite 3/4 of the available atoms to the upper state, corresponding to the ratio $15/20$ of the number of magnetic substates in the upper levels to the total. Current AO lasers are sufficiently powerful that the excitation fraction closely approaches this limit. However, only a fraction of the total number of sodium atoms are available for excitation because of limited spectral overlap as the laser linewidth is typically much smaller than the sodium linewidth. \textcolor{black}{This limitation might be overcome with further development of efficient broad-band modeless pulsed lasers \cite{Pique2003}.}

\vspace{1cm}

\stepcounter{table}
\tablefirsthead{\multicolumn{5}{L{8.4cm}}{\bfseries Table 2. Transitions for a monochromatic LGS having $\lambda \ge 300$ nm, $\lambda_\text{ex} \ge 150$ nm and $\log(\varepsilon) > 18.3$.  }  \\[6pt] }
\tablehead{\\[-7pt] \multicolumn{5}{l}{{\bfseries Table 2. continued}} \\  \hline} 
\tabletail{\\[-14pt]\midrule}
\begin{center}
\begin{supertabular}{lSSSS}
\hline
Species & {$\lambda_\text{ex}$} & {$\lambda$} &  {$ \log(\varepsilon)$}  & {$\log(N\sigma_{21})$} \\
              & {(nm)}            & {(nm)}   \\
\hline\\[-9pt]
Na I &  588.995 &  588.995 & 21.20 &  -1.50  \\ 
    &  589.592 &  589.592 & 21.07 &  -1.50  \\ 
    &  330.237 &  2205.647 & 20.00 &  -0.75  \\ 
    &  330.237 &  588.995 & 19.83 &  -1.50  \\ 
    &  330.298 &  2208.370 & 19.82 &  -0.75  \\ 
    &  330.237 &  1140.377 & 19.82 &  -1.19  \\ 
    &  330.298 &  588.995 & 19.65 &  -1.50  \\ 
    &  330.298 &  1140.377 & 19.65 &  -1.19  \\ 
    &  330.237 &  330.237 & 19.61 &  -3.61  \\ 
    &  330.237 &  589.592 & 19.52 &  -1.50  \\ 
    &  330.237 &  1138.144 & 19.52 &  -1.49  \\ 
    &  330.298 &  330.298 & 19.44 &  -3.61  \\ 
    &  330.298 &  589.592 & 19.37 &  -1.50  \\ 
    &  330.298 &  1138.144 & 19.35 &  -1.49  \\ 
    &  285.281 &  5426.880 & 19.20 &  -0.25  \\ 
    &  285.281 &  588.995 & 19.18 &  -1.50  \\ 
    &  285.301 &  5434.170 & 18.98 &  -0.25  \\ 
    &  285.301 &  588.995 & 18.95 &  -1.50  \\ 
    &  285.281 &  1140.377 & 18.93 &  -1.19  \\ 
    &  285.281 &  1074.645 & 18.91 &  -2.65  \\ 
    &  285.281 &  589.592 & 18.87 &  -1.50  \\ 
    &  285.281 &  616.075 & 18.78 &  -2.54  \\ 
    &  268.034 &  588.995 & 18.71 &  -1.50  \\ 
    &  285.301 &  1140.377 & 18.71 &  -1.19  \\ 
    &  285.301 &  1074.930 & 18.68 &  -2.65  \\ 
    &  285.301 &  589.592 & 18.67 &  -1.50  \\ 
    &  285.281 &  3414.261 & 18.64 &  -0.45  \\ 
    &  285.281 &  1138.144 & 18.63 &  -1.49  \\ 
    &  268.034 &  10807.960 & 18.61 &  0.14  \\ 
    &  285.301 &  616.075 & 18.56 &  -2.54  \\ 
    &  285.281 &  2205.647 & 18.51 &  -0.75  \\ 
    &  285.281 &  615.423 & 18.49 &  -2.84  \\ 
    &  268.043 &  588.995 & 18.45 &  -1.50  \\ 
    &  285.301 &  3414.261 & 18.42 &  -0.45  \\ 
    &  268.034 &  1140.377 & 18.41 &  -1.19  \\ 
    &  285.301 &  1138.144 & 18.41 &  -1.49  \\ 
    &  268.034 &  589.592 & 18.40 &  -1.50  \\ 
    &  268.043 &  10823.050 & 18.36 &  0.14  \\ 
    &  259.387 &  588.995 & 18.34 &  -1.50  \\ 
    &  285.281 &  3407.755 & 18.34 &  -0.75  \\ 
    &  330.237 &  9090.560 & 18.33 &  -0.57  \\ 
    &  330.237 &  819.482 & 18.33 &  -1.15  \\ 
    &  268.034 &  2440.182 & 18.32 &  -2.08  \\ 
    &  268.034 &  864.993 & 18.31 &  -3.44  \\   
Fe I &  371.993 &  371.993 &  20.40  &   -2.08   \\ 
    &  385.991 &  385.991 &  18.72  &   -2.26   \\ 
    &  344.061 &  344.061 &  18.49  &   -2.16   \\ 
    &  367.991 &  373.713 &  18.18  &   -2.14   \\ 
    &  296.690 &  373.486 &  18.17  &   -1.33   \\ 
    &  216.677 &  349.710 &  18.13  &   -2.42   \\ 
    &  302.064 &  302.064 &  18.13  &   -1.68   \\ 
    &  261.871 &  381.584 &  18.08  &   -1.21   \\ 
    &  302.064 &  382.042 &  18.07  &   -1.43   \\ 
    &  344.061 &  349.057 &  18.05  &   -2.59   \\ 
    &  213.202 &  340.746 &  18.04  &   -1.62   \\ 
    &  437.593 &  646.271 &  18.04  &   -3.59   \\ 
    &  319.166 &  516.749 &  18.03  &   -2.43   \\
Ca I &  422.673 &  422.673 & 19.72 &  -3.34  \\ 
K  I &  766.490 &  766.490 & 19.05 &  -3.21  \\ 
    &  769.896 &  769.896 & 18.92 &  -3.21  \\ 
Mg I &  202.582 &  1710.866 & 18.15 &  -1.55  \\ 
    &  202.582 &  1182.819 & 18.15 &  -1.63  \\[-1pt]
\hline
\end{supertabular}
  \label{tab:Return Flux MLGS}
\end{center}

Quantitative results for single-photon excitation in Na I are shown in Table 2. In Table 3 we list all transitions in Na I that are available with two-step excitation and have a return flux coefficient \textcolor{black}{$\varepsilon > 10^{20}$} photons s$^{-1}$ m$^{-2}$. There are many, but even the strongest is \textcolor{black}{two} times weaker that the sodium D$_2$ line. 

Of particular interest for PLGS are transitions from the 4P$_{3/2}$ state \cite{Pique2006}. This state may be populated directly from the ground state by absorption of a 330 nm photon. De-excitation occurs through two pathways. Approximately 29\% decay directly to the ground state, emitting a 330 nm photon and 70\% decay to the 4S$_{1/2}$ state, emitting a 2206 nm photon. That state can decay to the ground state via two pathways. About 67\% decay via the 3P$_{3/2}$ state, emitting 1140 and 589 nm photons. The remaining decay via the 3P$_{1/2}$ state emitting 1138 and 590 nm photons. About 1.5\% of the atoms in the 4P$_{3/2}$ state return to the ground state via the 3D$_{5/2}$ state, emitting 9093, 819 and 589 nm photons.

\stepcounter{table}
\tablefirsthead{\multicolumn{4}{L{8.4cm}}{\bfseries Table 3. Na I transitions for a monochromatic LGS employing two-step excitation having $\lambda \ge 300$ nm and $\log(\varepsilon) > 20$.   }  \\[6pt] } 
\tablehead{\\[-7pt] \multicolumn{4}{l}{{\bfseries Table 3. continued}} \\  \hline} 
\tabletail{\\[-14pt]\midrule}
\tablelasttail{}
\begin{center}
\begin{supertabular}{ScSS}
\hline
 {$\lambda$}     & $\lambda_\text{ex}$  &  {$\log(\varepsilon)$}  & {$\log(N\sigma_{21})$} \\ 
 {(nm)}                 & {(nm)}    \\
\hline
~\\[-9pt]
819.482 & 588.995+819.482 &   21.00  &  -1.15  \\ 
818.326 & 589.592+818.326 &   20.89  &  -1.23  \\ 
818.326 & 588.995+819.479 &   20.68  &  -1.23  \\ 
589.592 & 588.995+819.479 &   20.68  &  -1.50  \\ 
1140.377 & 589.592+1138.144 &   20.31  &  -1.19  \\ 
588.995 & 589.592+1138.144 &   20.31  &  -1.50  \\ 
1140.377 & 588.995+1140.377 &   20.22  &  -1.19  \\ 
568.820 & 588.995+568.820 &   20.19  &  -2.26  \\ 
819.479 & 589.592+818.326 &   20.19  &  -1.93  \\ 
588.995 & 589.592+818.326 &   20.19  &  -1.50  \\ 
588.995 & 330.237+2337.915 &   20.18  &  -1.50  \\ 
568.263 & 589.592+568.263 &   20.10  &  -2.34  \\ 
568.820 & 330.237+2337.915 &   20.08  &  -2.26  \\ 
1138.144 & 589.592+1138.144 &   20.01  &  -1.49  \\ 
\hline
\end{supertabular}
\end{center}

It is also possible to use a 330 nm photon to excite directly to the 4P$_{1/2}$ state, rather than the 4P$_{3/2}$ state, with a comparable transition rate. That state decays directly to the ground state with a 29\% probability, to the 4S$_{1/2}$ state with a 70\% probability, and to the 3D$_{3/2}$ state with a 1.6\% probability, producing a 9140 nm photon. The 3D$_{3/2}$ state decays to ground state via the 3P$_{1/2}$ state, producing 818 and 590 nm photons, or via the 3P$_{3/2}$ state, producing 819 and 589 nm photons. 

We searched for all possible PLGS line combinations employing single-photon excitation in Na I, with the condition that they have a return flux greater than $10^{17}$ photons s$^{-1}$ m$^{-2}$ (four orders of magnitude weaker than the Na D$_2$ lines) and a penalty factor less than 50. The results are shown in Table 4. The only possibilities are the two already mentioned. 

\tablefirsthead{\multicolumn{7}{L{8.3cm}}{\bfseries Table 4. Na I transitions for a PLGS. All transitions having $q > 0.4$ and $\lambda_\text{ex} \ge 300$ nm are listed. }  \\[6pt] }
\tablehead{\\[-7pt] \multicolumn{7}{l}{{\bfseries Table 4 continued}} \\  \hline} 
\tabletail{\\[-14pt]\midrule}
\tablelasttail{}
\begin{center}
\begin{supertabular}{SS @{\hspace{0.2 cm}} S @{\hspace{0.3 cm}} S @{\hspace{0.2 cm}} S @{\hspace{0.3 cm}} S @{\hspace{0.2 cm}} S}
\hline
{$\lambda_\text{ex}$}  & {$q$}   & {$p$}   & {$\lambda_{1}$} &  {$\log(\varepsilon_{1})$ }  & {$\lambda_{2}$} &  {$\log(\varepsilon_{2})$} \\
{(nm)}                             &           &         &     {(nm)}             &      &  {(nm)}      \\
\hline
~\\[-9pt]
330.237 &  1.00 &  18.59 &  330.237 &   19.61  & 2205.647 &  20.00  \\ 
 &  0.94 &  19.74 &  330.237 &  19.61 & 1140.377 &  19.82  \\ 
 &  0.84 &  19.74 &  330.237 &  19.61 & 1138.144 &  19.52  \\ 
 &  0.72 &  25.86 &  330.237 &  19.61 & 588.995 &  19.83  \\ 
 &  0.65 &  25.84 &  330.237 &  19.61 & 589.592 &  19.52  \\ 
\hline 
 ~\\[-9pt] 
330.298 &  0.82 &  18.60 &  330.298 &  19.44 & 2208.370 &  19.82  \\ 
 &  0.77 &  19.75 &  330.298 &  19.44 & 1140.377 &  19.65  \\ 
 &  0.69 &  19.75 &  330.298 &  19.44 & 1138.144 &  19.35  \\ 
 &  0.59 &  25.87 &  330.298 &  19.44 & 588.995 &  19.65  \\ 
 &  0.54 &  25.85 &  330.298 &  19.44 & 589.592 &  19.37  \\ 
\hline

\end{supertabular}
\end{center}

Several transitions involving two-step excitation have been discussed in the literature. The processes of 3S$_{1/2} \to $ 3P$_{3/2} \to $ 4D$_{5/2}$ ($589+569$ nm) \cite{Foy1995}, 3S$_{1/2} \to $ 3P$_{3/2} \to $ 4S$_{1/2}$ ($589 + 1140$ nm) \cite{Biegert2003} and 3S$_{1/2} \to $ 3P$_{3/2} \to $ 3D$_{5/2}$ ($589 + 819.7$ nm) \cite{Xu2017}. Excitation via the $3P_{1/2}$ state is also possible but with somewhat lower transition rates.  

Atoms excited to the 3P levels can be further excited to the 4D, 4S or 3D levels by absorption of visible or infrared light, provided that this is done within a few ns. Sodium can also be excited from the ground state to the 4D$_{5/2}$ level by absorption of two 578 nm photons. This is a non-linear process involving a virtual intermediate state, requiring picosecond pulses and a laser irradiance on the order of $10^{10}$ W m$^{-2}$ \cite{Biegert2003}.

All possible combinations employing two-step excitation in Na I are listed in Table 5. The selection criteria is $q > 0.4$. The strongest of these is $589 + 569$ nm excitation producing lines at 569 and 2334 nm. These transitions would provide a SNR that is 0.79 times the single-photon excitation transitions shown in Table 4.  That is roughly equivalent to a decrease to 65\% in LGS return flux. 

\stepcounter{table}
\tablefirsthead{\multicolumn{7}{L{8.3cm}}{\bfseries Table 5. Na I transitions for a PLGS employing two-step excitation. All transitions that have $q > 0.4$ and $\lambda_\text{ex} \ge 300$ nm are listed. The last line for each excitation scheme shows the return flux coefficient for the first excitation wavelength. }  \\[6pt] }
\tablehead{\\[-7pt] \multicolumn{7}{l}{{\bfseries Table 5 continued}} \\  \hline} 
\tabletail{\\[-14pt]\midrule}
\tablelasttail{}
\begin{center}
\begin{supertabular}{c @{\hspace{0.1 cm}} S @{\hspace{0.1 cm}} S @{\hspace{0.15 cm}} S @{\hspace{0.1 cm}} S @{\hspace{0.15 cm}} S @{\hspace{0.1 cm}} S}
\hline
{$\lambda_\text{ex}$}  & {$q$}   & {$p$}   & {$\lambda_{1}$} &  {$\log(\varepsilon_{1})$}         & {$\lambda_{2}$} &  {$\log(\varepsilon_{2})$} \\
{(nm)}                             &           &         &     {(nm)}             &      &  {(nm)}    \\
\hline
~\\[-9pt]
\hspace{1.9mm} 588.995 &  0.79  &   18.55 & 330.237  & 19.41 & 2337.915  & 19.95 \\ 
+ 568.820 &  0.79  &   18.59 & 330.237  & 19.41 & 2205.647  & 19.79 \\ 
 &  0.75  &   19.74 & 330.237  & 19.41 & 1140.377  & 19.62 \\ 
 &  0.67  &   19.74 & 330.237  & 19.41 & 1138.144  & 19.32 \\ 
 &  0.55  &   26.69 & 330.237  & 19.41 & 568.820  & 20.19 \\ 
 &  0.51  &   25.84 & 330.237  & 19.41 & 589.592  & 19.32 \\ 
 &  0.45  &   60.81 & 568.820  & 20.19 & 2337.915  & 19.95 \\ 
  &  &  & 588.995 &   20.78  \\ 
\hline 
 ~\\[-9pt] 
\hspace{1.9mm} 589.592 &  0.71  &   18.56 & 330.298  & 19.32 & 2334.843  & 19.86 \\ 
+ 568.263 &  0.71  &   18.60 & 330.298  & 19.32 & 2208.370  & 19.70 \\ 
 &  0.67  &   19.75 & 330.298  & 19.32 & 1140.377  & 19.60 \\ 
 &  0.66  &   19.75 & 330.298  & 19.32 & 1138.144  & 19.30 \\ 
 &  0.59  &   18.56 & 330.298  & 19.32 & 2337.896  & 19.16 \\ 
 &  0.51  &   25.87 & 330.298  & 19.32 & 588.995  & 19.82 \\ 
 &  0.49  &   18.60 & 330.298  & 19.32 & 2205.647  & 19.00 \\ 
 &  0.49  &   26.71 & 330.298  & 19.32 & 568.819  & 19.40 \\ 
 &  0.49  &   26.73 & 330.298  & 19.32 & 568.263  & 20.10 \\ 
 &  0.41  &   60.69 & 568.263  & 20.10 & 2334.843  & 19.86 \\ 
  &  &  & 589.592 &   20.65 \\ 
\hline 
 ~\\[-9pt] 
\hspace{1.9mm} 588.995 &  0.57  &   18.56 & 330.298  & 19.13 & 2334.843  & 19.67 \\ 
+ 568.819 &  0.57  &  18.60 & 330.298  & 19.13 & 2208.370  & 19.51 \\ 
 &  0.54  &   19.75 & 330.298  & 19.13 & 1140.377  & 19.41 \\ 
 &  0.53  &   19.75 & 330.298  & 19.13 & 1138.144  & 19.11 \\ 
 &  0.48  &   18.56 & 330.298  & 19.13 & 2337.896  & 18.97 \\ 
 &  0.41  &   25.85 & 330.298  & 19.13 & 589.592  & 19.98 \\ 
 &  0.40  &   18.60 & 330.298  & 19.13 & 2205.647  & 18.81 \\ 
 &  0.40  &   26.71 & 330.298  & 19.13 & 568.819  & 19.21 \\ 
 &  0.40  &   26.73 & 330.298  & 19.13 & 568.263  & 19.91 \\ 
   &  &  & 588.995 &   20.96 \\
\hline 
 ~\\[-9pt] 
\hspace{1.9mm} 330.237 &  0.51  &   18.56 & 330.298  & 19.03 & 2334.843  & 19.57 \\ 
+ 2337.896 &  0.51  &   18.60 & 330.298  & 19.03 & 2208.370  & 19.41 \\ 
 &  0.51  &   18.60 & 330.298  & 19.03 & 2205.647  & 19.64 \\ 
 &  0.48  &   19.75 & 330.298  & 19.03 & 1140.377  & 19.66 \\ 
 &  0.48  &   19.75 & 330.298  & 19.03 & 1138.144  & 19.36 \\ 
 &  0.43  &   18.56 & 330.298  & 19.03 & 2337.896  & 18.87 \\ 
 &  & & 330.237 &   19.25  \\
\hline 
 ~\\[-9pt] 
\hspace{1.9mm} 588.995 &  0.50  &   126.58 & 589.592  & 20.68 & 818.326  & 20.68 \\ 
+ 819.479 &  0.23  &   126.21 & 589.592  & 20.68 & 819.479  & 19.99 \\ 
  &  &  & 588.995 &   20.91  \\ 
\hline 
 ~\\[-9pt] 
\hspace{1.9mm} 589.592 &  0.50  &   83.39 & 588.995  & 20.31 & 1140.377  & 20.31 \\ 
+ 1138.144 &  0.35  &   83.50 & 588.995  & 20.31 & 1138.144  & 20.01 \\ 
  &  &  & 589.592 &   20.87  \\ 
\hline 
 ~\\[-9pt] 
\hspace{1.9mm} 588.995 &  0.49  &   18.27 & 330.237  & 18.99 & 5426.880  & 19.01 \\ 
+ 498.281 &  0.49  &   18.28 & 330.237  & 18.99 & 5013.920  & 19.30 \\ 
 &  0.49  &   18.59 & 330.237  & 18.99 & 2205.647  & 19.37 \\ 
 &  0.47  &   19.09 & 330.237  & 18.99 & 1477.974  & 19.49 \\ 
 &  0.46  &   19.74 & 330.237  & 18.99 & 1140.377  & 19.30 \\ 
 &  0.46  &   19.74 & 330.237  & 18.99 & 1138.144  & 18.99 \\ 
  &  &  & 588.995 &   20.78   \\ 
  \hline 
 ~\\[-9pt] 
\hspace{1.9mm} 589.592 &  0.47  &   18.37 & 330.237  & 18.94 & 3414.261  & 19.49 \\ 
+ 615.423 &  0.47  &   18.37 & 330.237  & 18.94 & 3407.755  & 19.19 \\ 
 &  0.46  &   18.59 & 330.237  & 18.94 & 2208.370  & 19.03 \\ 
 &  0.46  &   18.59 & 330.237  & 18.94 & 2205.647  & 19.33 \\ 
 &  0.44  &   19.74 & 330.237  & 18.94 & 1140.377  & 19.33 \\ 
 &  0.44  &   19.74 & 330.237  & 18.94 & 1138.144  & 19.03 \\ 
  &  &  & 589.592 &   20.86   \\ 
\hline 
 ~\\[-9pt] 
\hspace{1.9mm} 588.995 &  0.44  &   18.37 & 330.237  & 18.88 & 3414.261  & 19.42 \\ 
+ 616.075 &  0.44  &   18.37 & 330.237  & 18.88 & 3407.755  & 19.12 \\ 
 &  0.43  &   18.59 & 330.237  & 18.88 & 2208.370  & 18.96 \\ 
 &  0.43  &   18.59 & 330.237  & 18.88 & 2205.647  & 19.26 \\ 
 &  0.41  &   19.74 & 330.237  & 18.88 & 1140.377  & 19.26 \\ 
 &  0.40  &   19.74 & 330.237  & 18.88 & 1138.144  & 18.96 \\ 
  &  &  & 588.995 &   21.03  \\ 
\hline 
 ~\\[-9pt] 
\hspace{1.9mm} 589.592 &  0.44  &   18.28 & 330.298  & 18.88 & 5434.170  & 18.92 \\ 
+ 497.854 &  0.44  &   18.29 & 330.298  & 18.88 & 5007.670  & 19.21 \\ 
 &  0.43  &   18.60 & 330.298  & 18.88 & 2208.370  & 19.26 \\ 
 &  0.42  &   19.10 & 330.298  & 18.88 & 1476.749  & 19.40 \\ 
 &  0.40  &   19.75 & 330.298  & 18.88 & 1140.377  & 19.28 \\ 
 &  0.40  &   19.75 & 330.298  & 18.88 & 1138.144  & 18.98 \\ 
  &  &  & 589.592 &   20.66  \\
\hline 
 ~\\[-9pt] 
\hspace{1.9mm} 330.237 &  0.40  &   60.81 & 568.820  & 20.08 & 2337.915  & 19.85 \\ 
+ 2337.915   &  &  & 330.237 &   19.26   \\
\hline 
\multicolumn{7}{L{8.3cm}}{} \\
\end{supertabular}
\end{center}

\subsection{Iron}
The average iron abundances near the density peak at 90 km is two orders of magnitude higher than that of most other mesospheric metals, and twice that of sodium. Although its cross section is small, the key product $N\sigma_{21}$ is still larger than that of all other metallic elements except sodium. The abundance ratio Fe$^+/$Fe is approximately 0.2 above 90 km \cite{Feng2013}.

\begin{figure}[htbp!]
\centering
\includegraphics[width=6.4cm]{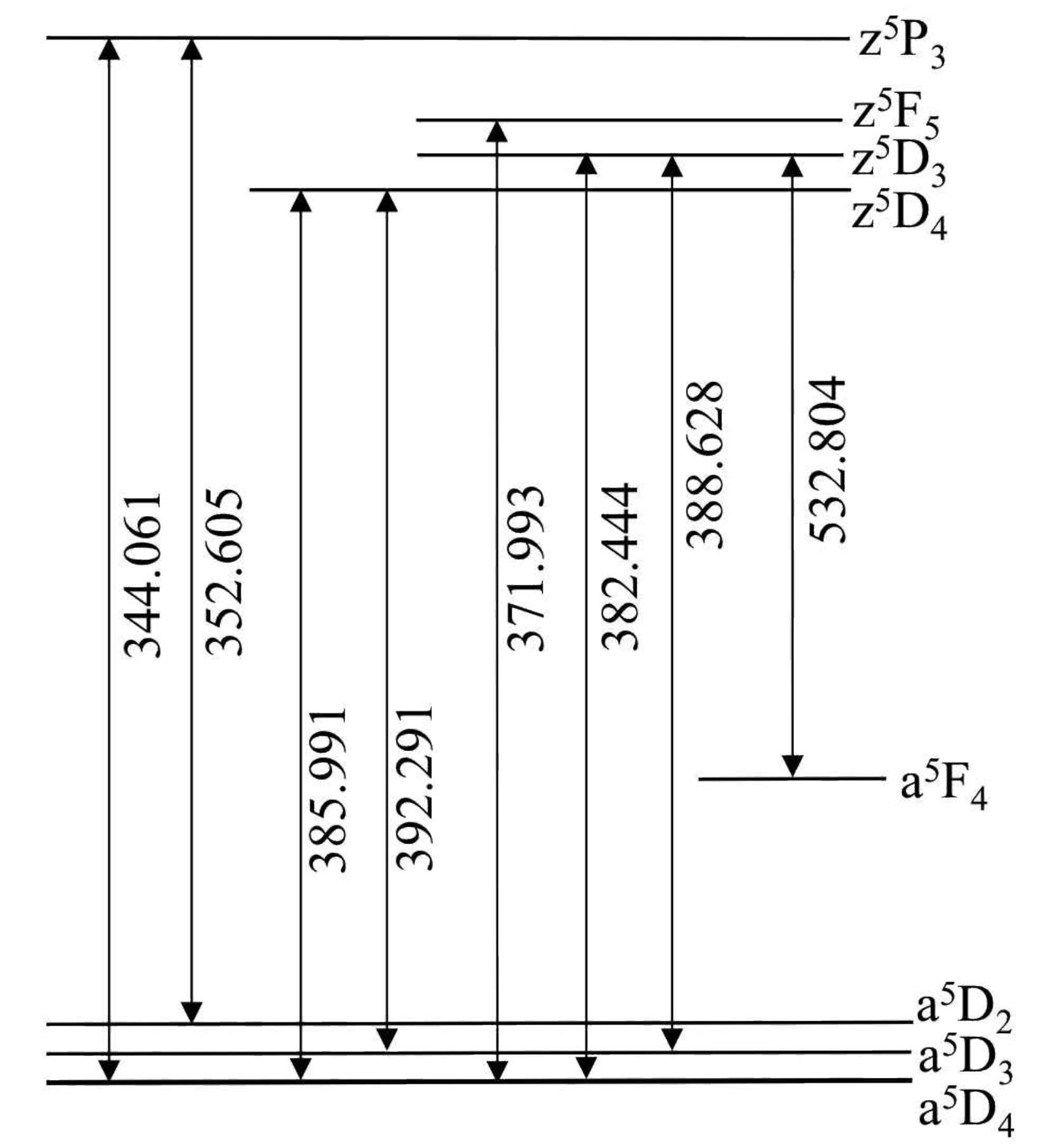}
\caption{Selected states and transitions of Fe I.}
\label{fig: fig02Fe}
\end{figure}

Atomic iron has the largest number of spectral lines of all mesospheric species. The iron-group elements have complex transitions with thousands of lines from the vacuum ultraviolet to the infrared. An energy-level diagram for Fe I is shown in Fig. \ref{fig: fig02Fe}. The ground state is split into three fine-structure levels. Only the lowest-energy a$^5$D$_4$ state is populated in thermal equilibrium in the mesosphere. There are five strong visible-wavelength transitions from this ground state, at 344.061, 367.991, 371.993, 382.444 and 385.991 nm. Most of these can decay to the other two ground states. The z$^5$D$_3$ state, reached by the 372 nm transition from the ground state, has a moderately-strong ($A = 1.15\times10^6$ s$^{-1}$) downward transition to the metastable a$^5$F$_4$ state. 

The strongest transition in Fe I using single-photon excitation has a return flux coefficient of \textcolor{black}{$2.51\times10^{20}$} photons s$^{-1}$ m$^{-2}$  (Table 2). This is \textcolor{black}{six} times weaker than the sodium D$_2$ line. Two-step excitation (Table 6) provides more possibilities, but these are all weaker by at least an order of magnitude. 

%
Fe I has combinations of transitions which have promising wavelengths and penalty factors for PLGS. However, for one-step and two-step excitation, $q$ does not exceed $ q > 0.18 $ for any combination of wavelengths (Tables 7 and 8).    

Fe II has no transitions from the ground state for wavelengths greater than 260 nm. This rules out excitation by ground-based lasers unless two-photon excitation is used. That would require pulsed lasers with very high irradiance.

\stepcounter{table}
\tablefirsthead{\multicolumn{4}{L{8.4cm}}{\bfseries Table 6. Fe I transitions for a monochromatic LGS employing two-step excitation. All transitions that have $\log(\varepsilon) > 19$ and $\lambda_\text{ex} \ge 300$ nm are listed.  }  \\[6pt] }
\tablehead{\\[-7pt] \multicolumn{4}{l}{{\bfseries Table 6 continued}} \\  \hline} 
\tabletail{\\[-14pt]\midrule}
\tablelasttail{}
\begin{center}
\begin{supertabular}{ccccc}
\hline
 {$\lambda$}     & $\lambda_\text{ex}$  &  $\log(\varepsilon)$  & $\log(N\sigma_{21})$ \\ 
 {(nm)}                 & (nm)               &  &  \\
\hline
~\\[-9pt]
355.492 & 437.593+355.492 & $  19.32 $ & $ -1.21 $ \\ 
314.399 & 385.991+314.399 & $  19.31 $ & $ -1.73 $ \\ 
381.764 & 371.993+381.764 & $  19.29 $ & $ -2.37 $ \\ 
380.198 & 371.993+380.198 & $  19.08 $ & $ -2.70 $ \\
\hline
\end{supertabular}
  \label{tab:Return Flux MLGS two photon Fe}
\end{center}

\tablefirsthead{\multicolumn{7}{L{8.3cm}}{\bfseries Table 7. Fe I transitions for a PLGS employing one-step excitation. All transitions that have $q > 0.1$ and $\lambda_\text{ex} \ge 300$ nm are listed. }  \\[6pt] }
\tablehead{\\[-7pt] \multicolumn{7}{l}{{\bfseries Table 7 continued}} \\  \hline} 
\tabletail{\\[-14pt]\midrule}
\tablelasttail{}
\begin{center}
\begin{supertabular}{cc @{\hspace{0.2 cm}} c @{\hspace{0.3 cm}} c @{\hspace{0.2 cm}} c @{\hspace{0.3 cm}} c @{\hspace{0.2 cm}} c}
\hline
{$\lambda_\text{ex}$}  & $q$   & $p$    & $\lambda_{1}$ &  $\log(\varepsilon_{1})$         & $\lambda_{2}$ &  $\log(\varepsilon_{2})$ \\
(nm)                             &           &         &     (nm)             &       &  (nm)      &    \\
\hline
~\\[-9pt]
319.323 &  0.10 &  26.39 &  319.323 &  $ 17.90 $ & 517.160 & $  17.90 $  \\
\hline
\end{supertabular}
\end{center}

\subsection{Magnesium}

Magnesium is one of the more abundant meteoritic constituents, 9.6\% by mass, so that meteoric ablation should inject large quantities of this metal into the MLT region \cite{Whalley2011}. Compared to other meteoric metals, magnesium has the largest ionization fraction, with a ratio Mg$^+$/Mg  in the range 4 to 12 \cite{Whalley2011}. 

Mg I has six strong transitions from the ground state, all in the vacuum ultraviolet. The strongest of these goes to the 3$^1$P$_1$ state, with a wavelength of  285.213 nm and a rate ($A_{21} = 4.91\times10^8$ s$^{-1}$) that is an order of magnitude greater than that of the sodium D lines. The only other downward transitions from this upper state are forbidden lines that have negligible rates. In vacuum, this would be a strong candidate for LGS, however, the wavelength is beyond the atmospheric cutoff. 

\stepcounter{table}
\tablefirsthead{\multicolumn{7}{L{8.3cm}}{\bfseries Table 8. Fe I transitions for a PLGS employing two-step excitation. All transitions that have a de-excitation $q > 0.1$ and $\lambda_\text{ex} \ge 300$ nm are listed. In brackets in the last line for each excitation scheme the return flux coefficient for the first excitation wavelength is shown. }  \\[6pt] }
\tablehead{\\[-7pt] \multicolumn{7}{l}{{\bfseries Table 5 continued}} \\  \hline} 
\tabletail{\\[-14pt]\midrule}
\tablelasttail{}
\begin{center}
\begin{supertabular}{c @{\hspace{0.1 cm}} S @{\hspace{0.1 cm}} S @{\hspace{0.15 cm}} S @{\hspace{0.1 cm}} S @{\hspace{0.15 cm}} S @{\hspace{0.1 cm}} S}
\hline
{$\lambda_\text{ex}$}  & {$q$}   & {$p$}   & {$\lambda_{1}$} &  {$\log(\varepsilon_{1})$}         & {$\lambda_{2}$} &  {$\log(\varepsilon_{2})$} \\
{(nm)}                             &           &         &     {(nm)}             &      &  {(nm)}    \\
\hline
 ~\\[-9pt] 
\hspace{1.9mm} 437.593 &  0.12  &   25.41 & 361.207  & 18.17 & 973.857  & 18.01 \\ 
+ 361.207 &  0.11  & 27.33 & 371.993  & 18.85 & 973.857  & 18.01 \\ 
  &  &  & 437.593 &   16.06  \\ 
\hline 
 ~\\[-9pt] 
\hspace{1.9mm} 385.991 &  0.12  &   40.78 & 371.993  & 18.49 & 561.564  & 18.49 \\ 
+ 532.418 &  0.10  &   44.57 & 371.993  & 18.49 & 532.418  & 18.38 \\ 
  &  &  & 385.991 &   18.18  \\ 
\hline 
\multicolumn{7}{L{8.3cm}}{} \\
\end{supertabular}
\end{center}

\begin{figure}[htbp!]
\centering
\includegraphics[width=7.6cm]{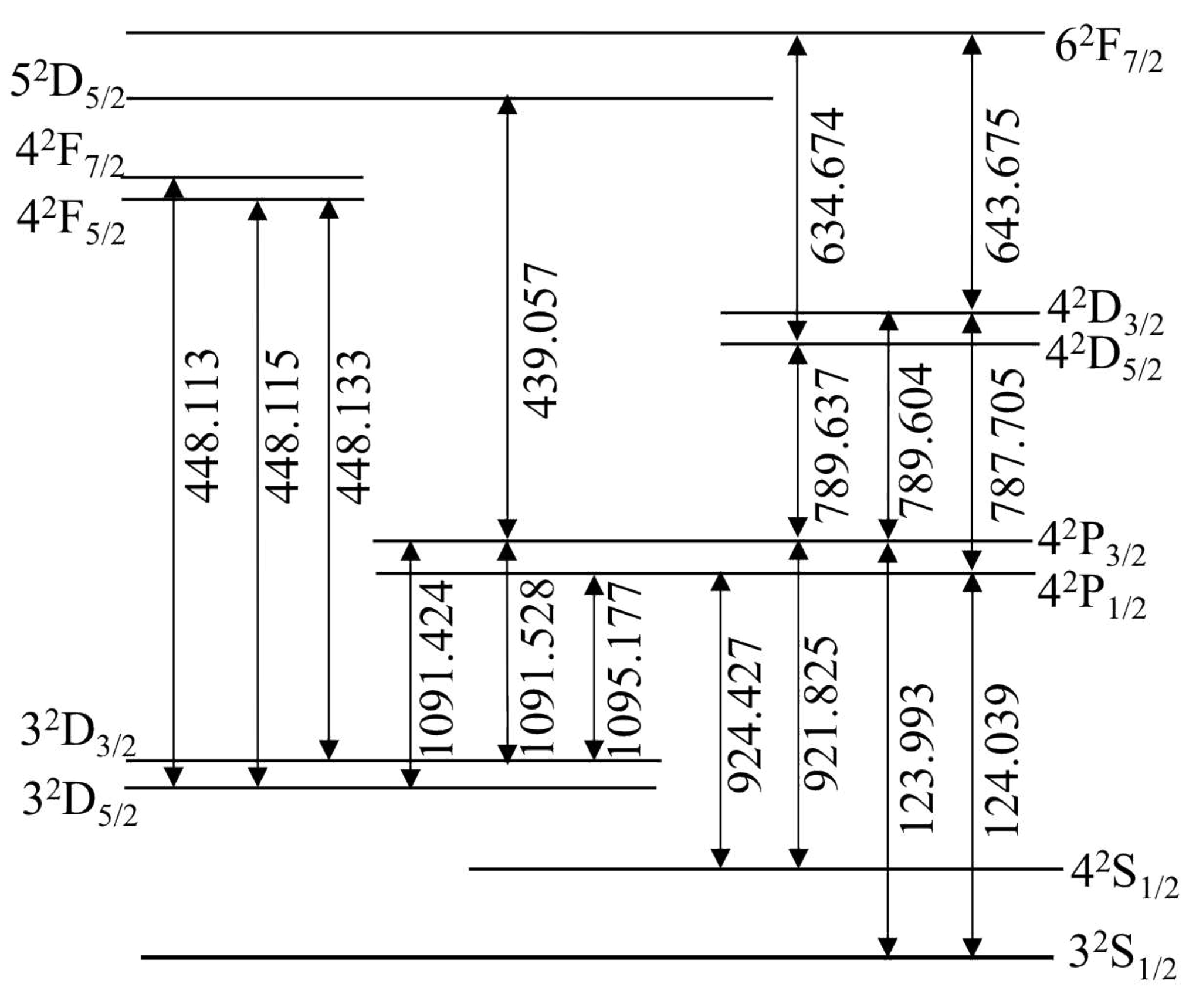}
\caption{Selected states and transitions of Mg II.}
\label{fig: fig03Mg+}
\end{figure}

Mg II has two transitions from the ground state that have rates greater than $10^8$ s$^{-1}$. These go to the 3$^2$P$_{1/2}$ and 3$^2$P$_{3/2}$ states, with wavelengths of 280.271 and 279.553 nm respectively. Both of these upper states return directly to the ground state, with no other significant transitions. Given the high abundance of Mg$^+$ (twice that of Na), these would be very good candidates for LGS, but they are beyond the atmospheric cutoff. 

Mg II does have strong emission lines in the visible region \cite{Kelleher2008a} (Fig.\ref{fig: fig03Mg+}). However, populating these levels from the ground state requires photons having wavelengths of 124 nm or less. This probably rules out Mg$^+$ for ground-based adaptive optics.

\subsection{Silicon}

Silicon is one of the most abundant elements in cosmic dust, comprising around 11\% by mass, and meteoric ablation injects a significant amount of Si into the atmosphere \cite{Plane2016}. However, neutral Si is oxidized to SiO very rapidly by reaction with O$_2$ at the altitude of maximum ablation, around 90 km. Si$^+$, SiO, and Si(OH)$_4$ are predicted to be the major silicon species above 80 km. Below 97 km the dominant sink is Si(OH)$_4$. The column density of Si$^+$ varies seasonally by an order of magnitude and peaks at $4.0\times 10^9$ cm$^{-2}$ during the summer \cite{Plane2016}.

\begin{figure}[htbp!]
\centering
\includegraphics[width=9cm]{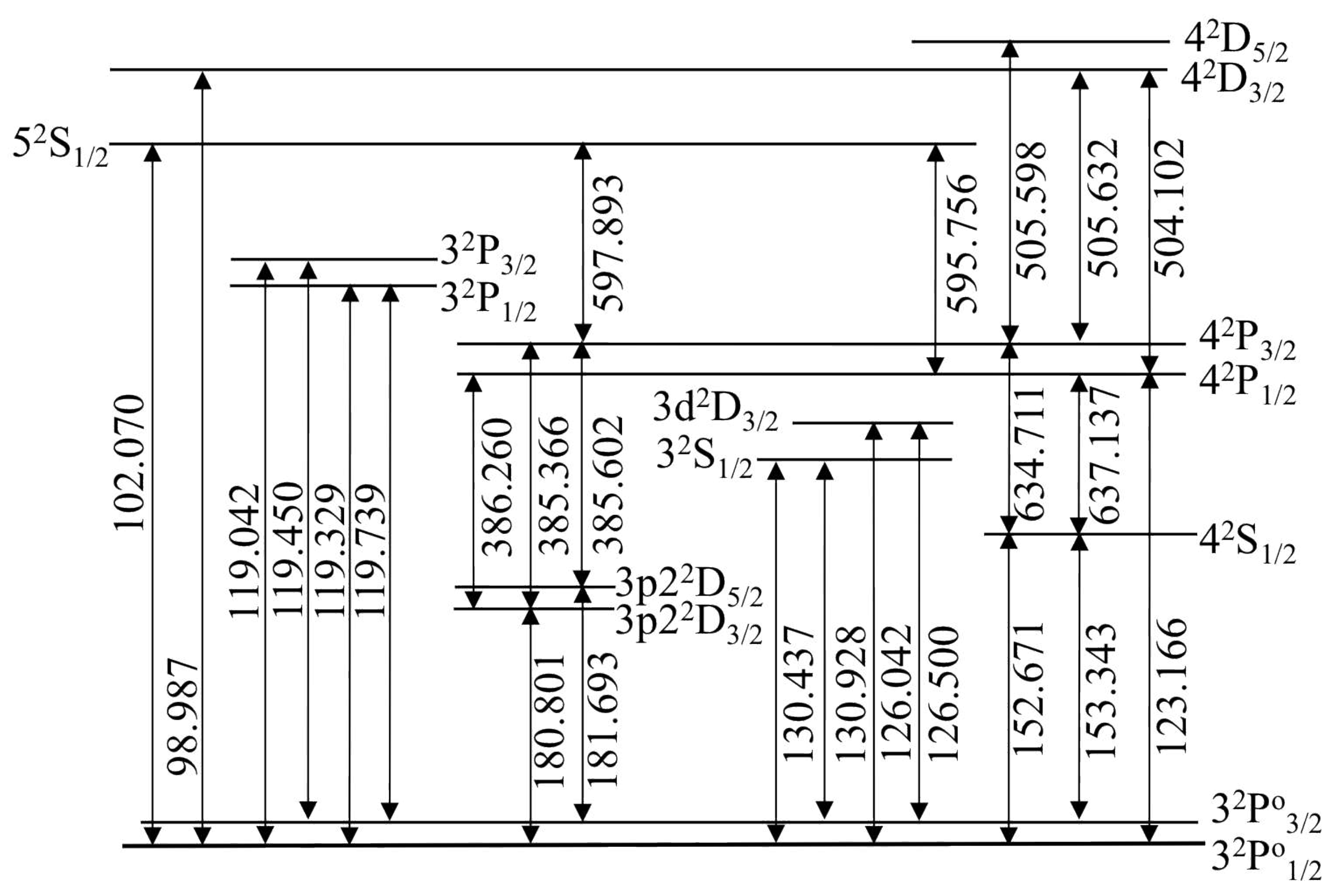}
\caption{Selected states and transitions of Si II.}
\label{fig: fig04Si+}
\end{figure}

Singly-ionized Silicon has several strong visible-light transitions, as shown in Fig. \ref{fig: fig04Si+} \cite{Kelleher2008b}. However, excitation from the ground state requires UV photons. The ground state is a doublet (3$^2$P$_{1/2}$,  3$^2$P$_{3/2}$), separated by 0.0356 eV. This is comparable to the mean particle energy of $3kT/2 \simeq 0.023$ eV, so both states will be populated. Transitions can occur from the lower ground state to the 4P$_{1/2}$ state, followed by a 637 nm visible-light transition, but this requires a 123-nm photon. The lowest-energy transitions from either ground state to the 4P levels requires a 123-nm photon. There is a transition from the ground state to the 3$^2$P$_{3/2}$ state, but this is quite weak ($A_{21} = 2.54\times10^6$ s$^{-1}$). This likely rules out SI II as a viable candidate for AO, and indeed no transitions met our thresholds for inclusion. 

\subsection{Calcium}
Lidar measurements of calcium showed a peak density of $2\times10^7$ atoms m$^{-3}$, about 200 times lower than the typical sodium atom density and 400 times lower than that of iron \cite{Plane2011}. The annual average Ca$^+$/Ca column density ratio is 2.4, the second largest ratio after magnesium. In comparison, the ratios of Na$^+$/Na and Fe$^+$/Fe are only about 0.2 \cite{Kopp1997}. Ca$^+$ is the dominant form above 90 km altitude, peaking near 105 km. 

Ca I has a strong transition from the ground state to the 3$^1$P$_1$ state with a rate of $2.18\times10^8$ s$^{-1}$ and a wavelength of 422.673 nm. The upper state returns directly to the ground state, with no other transitions. The only problem here is the low abundance. As can be seen in Table 2, the return flux coefficient is about \textcolor{black}{30} times smaller than that of the sodium D$_2$ line.

\stepcounter{table}
\tablefirsthead{\multicolumn{4}{L{8.4cm}}{\bfseries Table 9. Ca I transitions for a monochromatic LGS employing two-step excitation having wavelength $\lambda \ge 300$ nm and return-flux coefficient $\log(\varepsilon) > 19$.   }  \\[6pt] } 
\tablehead{\\[-7pt] \multicolumn{4}{l}{{\bfseries Table 9. continued}} \\  \hline} 
\tabletail{\\[-14pt]\midrule}
\tablelasttail{}
\begin{center}
\begin{supertabular}{Sccc}
\hline
 {$\lambda$}     & $\lambda_\text{ex}$  &  $\log(\varepsilon)$  & $\log(N\sigma_{21})$ \\ 
 {(nm)}                 & (nm)               &  &  \\
\hline
~\\[-9pt]
585.745 & 422.673+585.745 &  19.07  & -3.43  \\ 
430.774 & 657.278+430.774 &  19.01  & -3.35  \\
\hline
\end{supertabular}
  \label{tab:Return Flux MLGS two photon CaI}
\end{center}

\begin{figure}[htbp!]
\centering
\includegraphics[width=7.5cm]{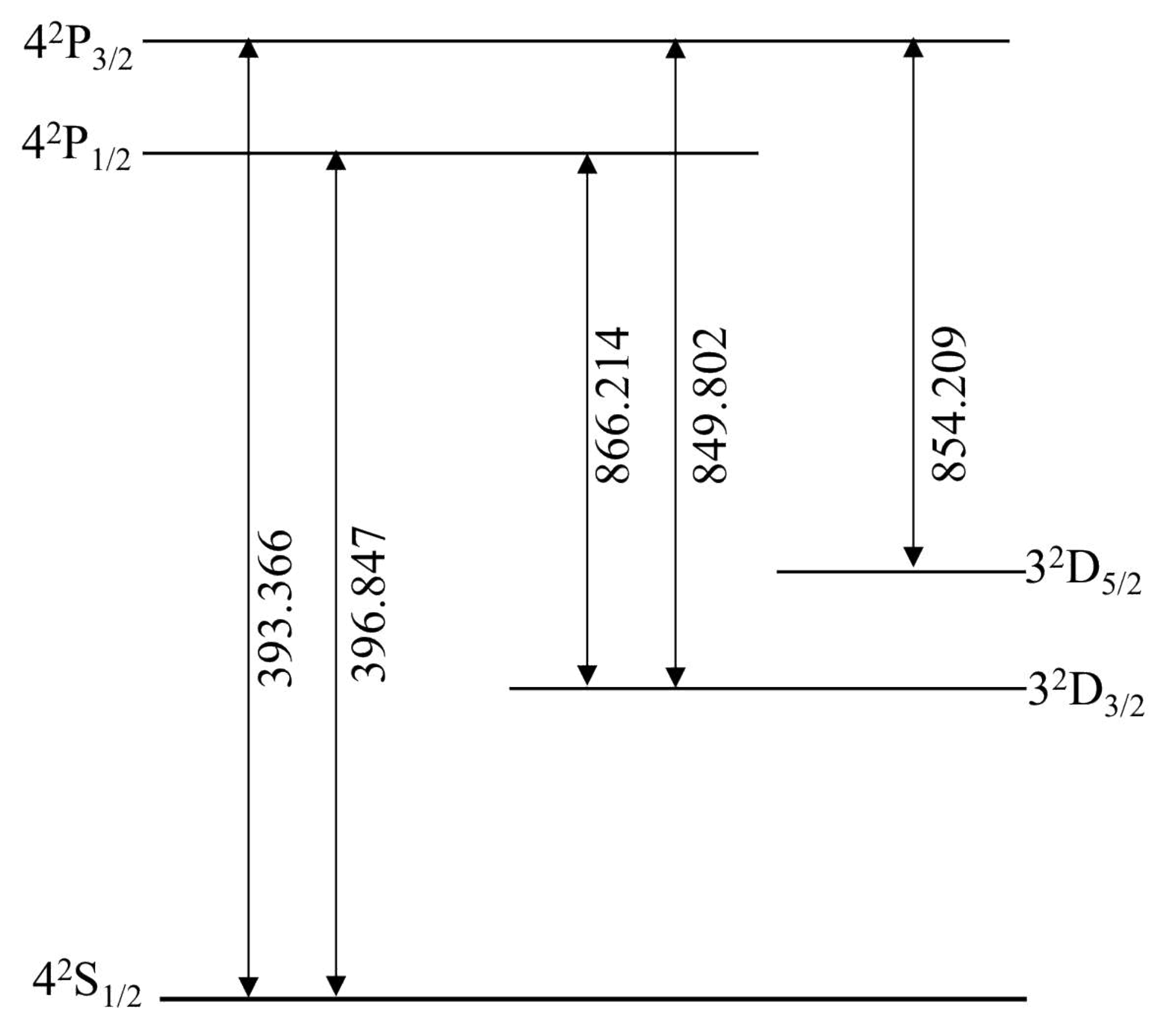}
\caption{Selected states and transitions of Ca II.}
\label{fig: fig05Ca+}
\end{figure}

Fig. \ref{fig: fig05Ca+} shows the relevant energy levels and transitions of  Ca$^+$ \cite{Maurer2004}. Excitation from the ground state to the 4P$_{1/2}$ and 4P$_{3/2}$ levels is possible using 397 nm and 393 nm photons, respectively. These states decay quickly to the ground state and also, with 7\% probability, to the 3D$_{3/2}$ and 3D$_{5/2}$ states, emitting 866 nm, 850 nm and 854 nm photons.
These states are metastable, so most of the atoms excited from the ground state using a UV laser will become trapped in these states. No lines of Ca II met the criteria for inclusion in Table 2. 

\subsection{Potassium}
\begin{figure}[htp!]
\centering
\includegraphics[width=8cm]{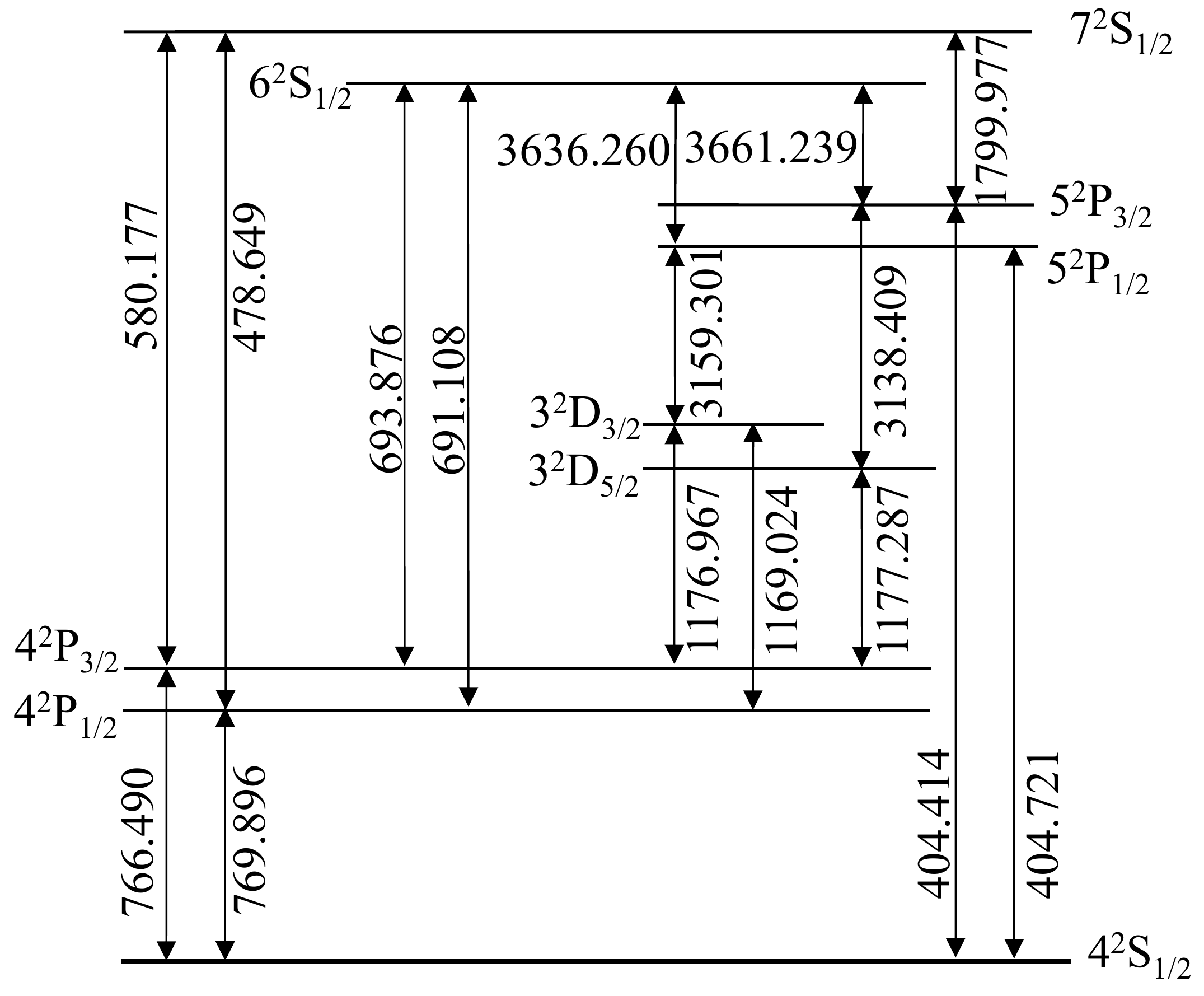}
\caption{Selected states and transitions of K I.}
\label{fig: fig06K}
\end{figure}

Potassium has relatively low abundance in the upper atmosphere, having a column density that is about two orders of magnitude less than that of Na. For neutral potassium atoms, many transitions covering the visible light range have high transition probabilities. The relevant energy levels and transitions are shown as Fig. \ref{fig: fig06K}. Its persistent lines are 769.896, 766.490, 404.414 and 404.721 nm. Other strong lines having slightly smaller transition probabilities  occur at higher energy levels.  Direct transitions to and from the ground state are possible at 766.490 and 769.896  nm. These have cross sections that are comparable to that of the sodium D$_2$ line. However, the relatively low column density of potassium results in return flux coefficients that are two orders of magnitude smaller.

\begin{figure*}[htb!]
\centering
\includegraphics[width=12.5cm]{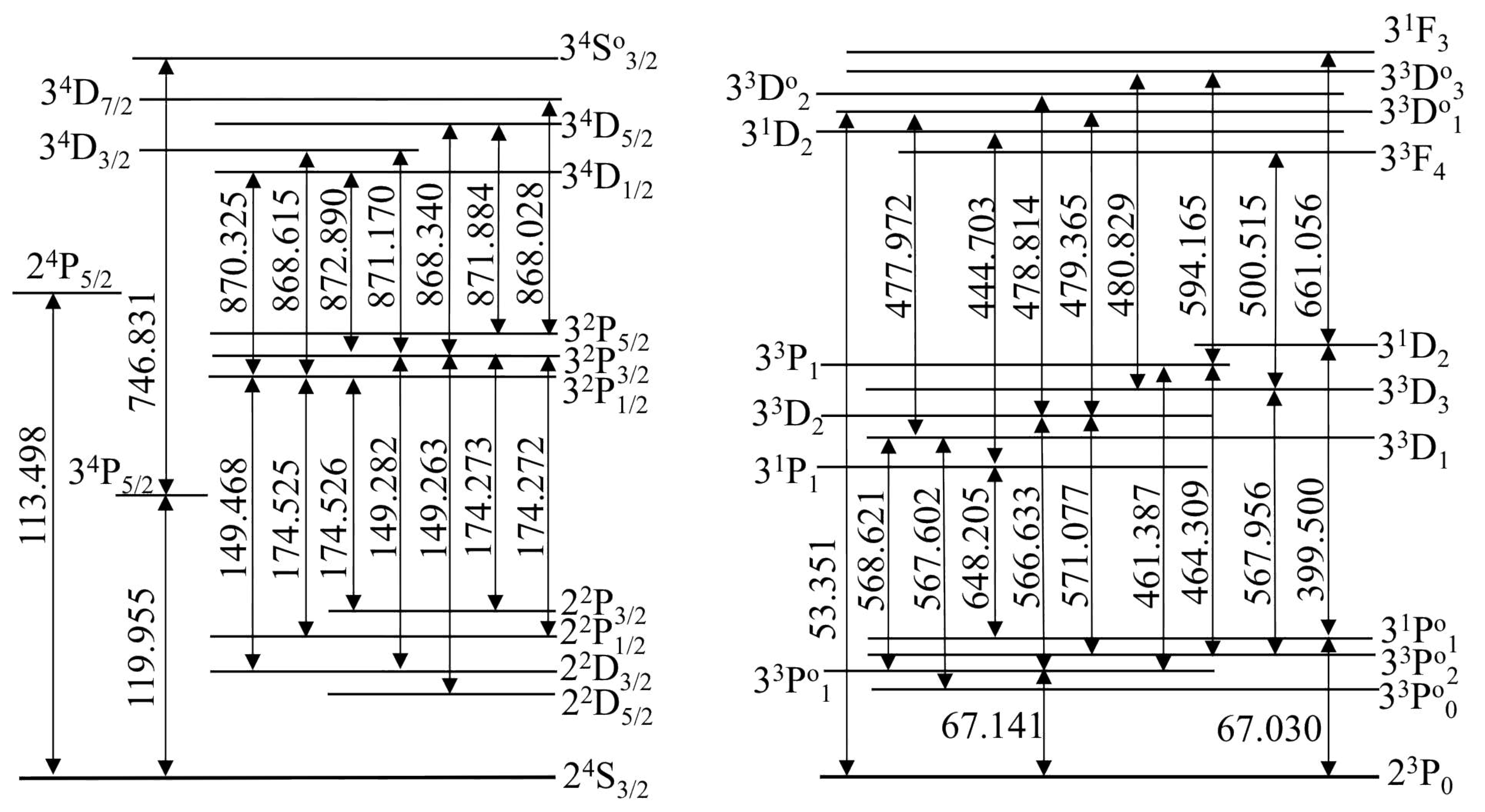}
\caption{Energy level diagram of N I (left) and N II (right).}
\label{fig: fig07N}
\end{figure*}

Potassium can also be excited directly from the ground state to the 5P$_{1/2}$ and 5P$_{3/2}$ levels. These have a path back to the ground state via the 3D states and 4P states, producing strong lines at 3139.265, 3160.163, 1169.024 and 1176.967 nm, in addition to the 766.490 and 769.896 nm doublet. This would be an attractive system for PLGS if it was not for the relatively low abundance.

\subsection{Nitrogen}

In the mesosphere, nitrogen molecules are ionized and dissociated by energetic electrons and protons,  producing N$_2^+$ by direct ionization and N$^+$ by dissociative ionization. Secondary electrons associated with ionization produce N in both the ground and 2D states. At 80 km atomic nitrogen has a peak density of $2 \times 10^{8}$ cm$^{-3}$,  falling to $10^6$ cm$^{-3}$ at 70 km \cite{Aikin2000}. The estimated column density of atomic nitrogen is $\sim 1 \times 10^{14}$ cm$^{-2}$.

The most relevant energy levels and transitions of atomic NI and singly-ionized NII are shown in Fig.\ref{fig: fig07N}. It can be seen that NII has many more strong emission lines in the visible range than does NI. Furthermore, nitrogen-discharge experiments \cite{Conti2001} show that the optical emission spectrum in the visible and UV regions, is quite complex with multiple-peaks and many blended lines. 

Direct transitions from the ground state require vacuum-ultraviolet photons, even using two-photon excitation. This likely rules out  nitrogen for ground-based AO systems. 

\subsection{Oxygen}

Atomic oxygen (O) is a fundamental component in chemical aeronomy of Earth's MLT region extending from approximately 50 km to over 100 km altitude. Primarily, O is generated through photolysis of molecular oxygen by UV radiation. The peak atomic oxygen concentrations, $\sim 6.0 - 6.5\times 10^{17}$ m$^{-3}$, are found at an altitude of approximately 95 km \cite{Mlynczak2013}. Assuming an atomic oxygen column width of 10 km in the mesosphere, the estimated column density of atomic oxygen is $\sim 6.0 - 6.5\times 10^{21}$m$^{-2}$.

\begin{figure*}[htbp!]
\centering
\includegraphics[width=12.5cm]{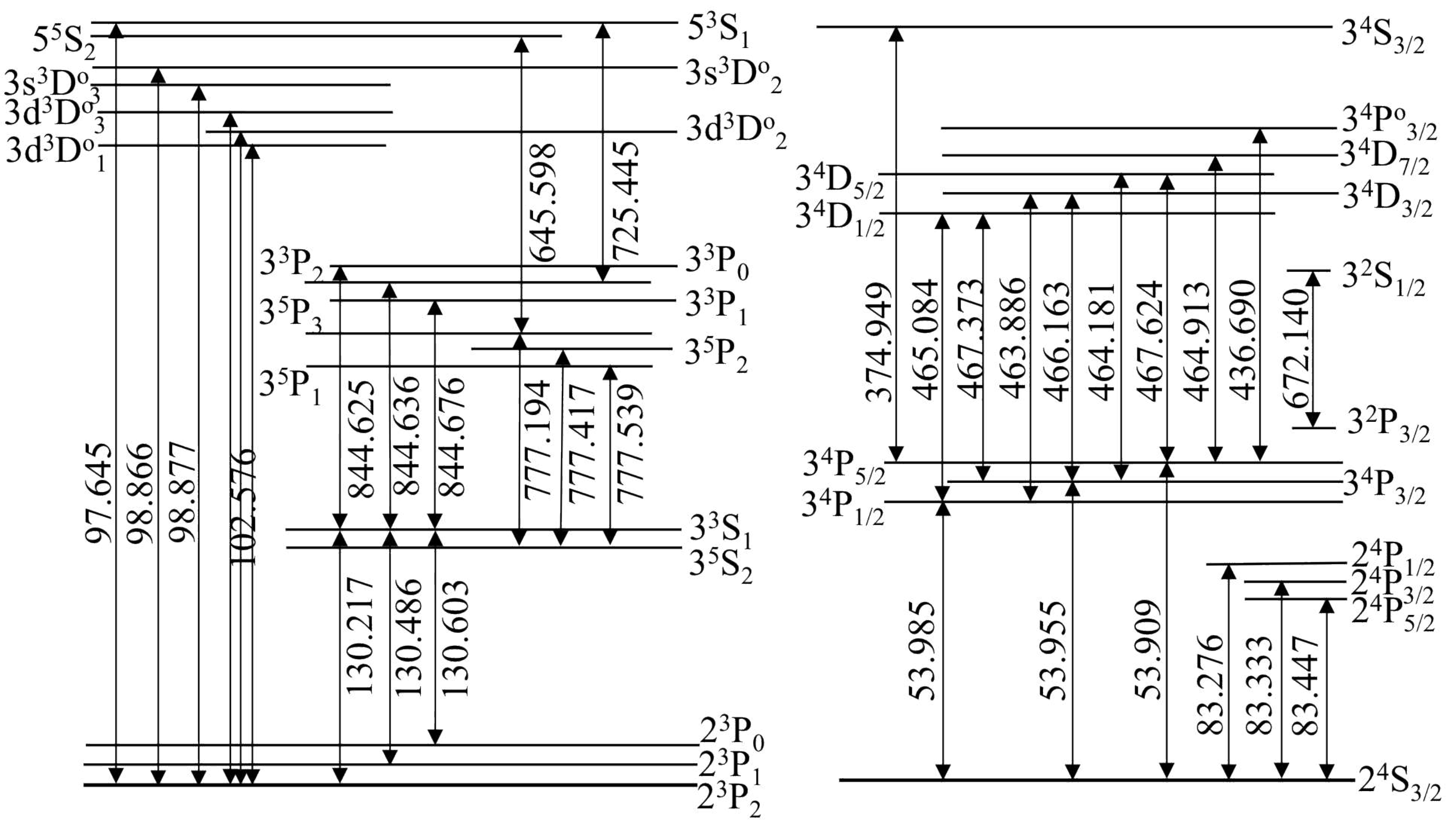}
\caption{Strong resonance transitions in O I (left) and O II (right).}
\label{fig: fig08O}
\end{figure*}

Energy level diagrams for O and O$^+$ are shown in Fig. \ref{fig: fig08O}. The triplet at $\sim 130$ nm has been observed in the cometary spectra from above the Earth's atmosphere \cite{Cochran2001}. These lines arise when O atoms in the ground state are excited to the 5S$_1$ state by solar photons. In the visible spectrum region, several astrophysically-important forbidden transitions exist. For O, these are the red doublet at 630.030 nm and 636.377 nm, arising from transitions between the 2D$_2$ and the 2P$_2$ and the 2P$_1$ levels respectively, and the green line at 557.734 nm from the 2D$_2$ and 2S$_0$ levels. These are magnetic dipole and electric quadrupole transitions respectively. These forbidden oxygen emission lines likely arise from atoms produced directly in the excited 2S$_0$ or 2D$_2$ states by photo-dissociation of parent molecules such as H$_2$O, CO and CO$_2$. In O$^+$, the doublet at 372.602 and 372.882 nm arises from transitions from the 2D$_{3/2}$ and 2D$_{5/2}$ levels to the 2S$_{3/2}$ ground state. As with nitrogen, $N\sigma_{21}$ is high for permitted transitions in oxygen, but low for the forbidden transitions.

In O, strong visible lines arise from transitions between the 3P and 3S levels, and also the 5S$_1$ level. Visible lines can also occur from states above the 5S$_1$ energy level. Each line is a doublet or triplet due to hyperfine structure. However, all these lines require vacuum-ultraviolet photons for excitation from the ground state. 

In O$^+$, strong visible lines result from transitions from the 3P and 3D levels to the 3P levels, and also the 3S$_{1/2}$ and 3P$_{3/2}$ levels. In this case there are no direct transitions from the ground state to the upper states involved, and transitions to the lower states require photons of wavelength 54 nm or less.

\subsection{Hydrogen}

Atomic hydrogen (H) is also a fundamental component in the photochemistry and energy balance of the mesopause region between approximately 80 and 100 km altitude. H is generated primarily by photolysis of water vapor and participates in a highly-exothermic reaction with ozone. During the day, atomic hydrogen concentrations peak in excess of $2.25\times 10^{14}$ m$^{-3}$ at higher latitudes near 85 km \cite{Mlynczak2014}. The estimated column density of atomic hydrogen is $\sim 2\times 10^{18}$ m$^{-2}$.

Hydrogen and hydrogen-like ions have familiar Lyman, Balmer, and Paschen spectral series. Hydrogen emission spectral lines in the visible range include Balmer lines at 410.173, 434.047, 486.134, and 656.283 nm wavelengths,  
corresponding to transitions to the $n = 2$ level. All energy levels for a given principal quantum number $n$ are degenerate. Transitions from the ground state require vacuum-ultraviolet photons having wavelengths of 122 nm or less.

\subsection{Carbon}

Meteoric ablation injects a significant amount of carbon into the mesosphere. However, atomic carbon is a very short-lived species.  It is stabilized in various multi-atomic structures having different molecular configurations. The carbon monoxide (CO) mixing ratio increases steeply with altitude in the mesosphere and in the lower part of the thermosphere is a result of the photolysis of carbon dioxide (CO$_2$) \cite{Grossmann2006}. Additionally, as the chemical lifetime of CO is much longer than the characteristic dynamical times, this gas is an excellent tracer of dynamics due to its strong 4.7 $\mu$m transition.

The most common oxidation state of carbon in inorganic compounds is $+4$, while $+2$ is found in carbon monoxide. The four outer electrons of carbon are valence electrons, so its ionization energies are much higher than those of most other elements. Short wavelength UV photons are needed to excite the atoms from the ground state, making excitation using a ground-based laser unfeasible. Also, there is little information about the detailed distribution and abundance of carbon and their ions in mesosphere, so carbon and related species are not considered further here. 

\section{Discussion}
\label{sec:discussion}

It can be seen from Table 2, that the sodium $D_2$ line provides the highest return flux of all possible transitions, including two-step excitation schemes. This is due to its relatively-large natural abundance and large cross section compared to other metal atoms. Mg$^+$ and Si$^+$ have stronger lines, but lack an effective scheme for excitation from the ground state. Although  the column density of Fe is twice that of sodium, its lower cross-section results in a return flux that is \textcolor{black}{one to two} orders of magnitude smaller. 

In contrast, the Mg$^+$ lines at 280.271 and 279.553 nm have transition rates that are an order of magnitude greater than that of the sodium D lines, and the column density of Mg$^+$ is twice that of sodium. These would be very strong lines if they were not beyond the atmospheric cutoff. 
Similarly,  Si$^+$ has strong transitions, and a column density comparable to that of Na, but vacuum-ultraviolet excitation is required. 

Potassium, like Ca$^+$, has a column density that is smaller than that of sodium by two orders of magnitude, so its return flux is small.

As with monochromatic LGS, Na yields the highest performance for PLGS from both one-step and two-step excitation. According to our results, the next-ranked element, Fe, provides at most \textcolor{black}{10\%} of the SNR of the best-suited transitions in Na. If atmospheric absorption can be compensated by sufficiently-high laser power the one-step excitation at 330 nm yields a higher SNR than the two-step excitation at $569+589$ nm. At an altitude of 4200 m, the atmospheric transmission at 330 nm is about 60\% of the transmission at 600 nm. The most-promising combinations of transitions for PLGS include de-excitations at 330 nm, however atmospheric absorption will decrease the return flux at 330 nm. For \textcolor{black}{two-step} excitation, an alternative approach could be to combine the return flux from the D$_{1}$ and D$_{2}$ line at 589 nm. For an excitation at $589 + 569$ nm, the photons at 589 nm can only be used if it was ensured that the photons were emitted from fully-coincident laser spots in the mesosphere. Even though the two-step excitation adds wavelengths at about 3410 and 2338 nm, the PLGS performance does not exceed that of the one-step excitation, since the penalty factor does not decrease significantly for infrared wavelengths.    

Our analysis used statistical weights provided by \textcolor{black}{NIST}, which give the number of hyperfine structure states. Magnetic sub-states arising from Zeeman splitting in an external magnetic field are not considered. This is a limitation to our approach. Excitation with optimized laser polarization, which can provide a greater return flux, can be assessed by employing a more-detailed treatment of single transitions using Bloch-equations. Nevertheless, we believe that the simple approach that we employ is useful for a first assessment of potential transitions. 

For a species in which essentially all atoms are in the ground state, \textcolor{black}{the} optical depth at the line center is equal to the product of column density, $\tau_0 = N\sigma_{12}(\nu_0) = g_2 N\sigma_{21}(\nu_0)/g_1$. From the values listed in Table C1, we see that all transitions of mesospheric metals have optical depth less than 0.2, so the atmosphere is optically thin ($\tau < 1$). In contrast, the abundant species of N I and O I have $\tau_0 \gg 1$ for most permitted transitions, which means that the atmosphere is opaque at those wavelengths. 

The optical depth for stimulated emission is closely related to the optical depth for absorption. In a fully-excited medium, $\tau \sim - \sigma_{21}N$. Significant amplification by stimulated emission can only occur in a medium for which $-\tau \gg 1$. To achieve an optical depth of  order unity, $N_2A_{21}\geq 10^{23}$~m$^{-2}$s$^{-1}$ is required. For a typical transition rate, $A_{21}\sim 10^{7}$~s$^{-1}$, the column density of the medium must therefore exceed $10^{16}$~m$^{-2}$ to achieve ASE. This would appear to rule out all metallic species in the mesosphere.

An interesting new development is the use of high-powered lasers to create a plasma in the atmosphere. Rairoux et al. \cite{Rairoux2000} have demonstrated ASE in backscattered fluorescence of nitrogen from filaments generated by intense ultra-fast Ti:sapphire laser pulses propagating over a distance of up to several km. As it propagates along the filament, the backscattered radiation is amplified by a population inversion resulting from direct ionization. The laser excitation leads to plasma formation and the consequence is a white-light laser pulse. Backscattered fluorescence from N$_2$ molecules and ions shows the exponential increase with increasing filament length expected for ASE. It has the potential to generate very bright LGS, but a terawatt femtosecond Ti:sapphire laser system is required to produce the $10^{17}$ W/m$^2$ intensity that is necessary to generate non-linear effects and plasma. Nevertheless, the potential of a white-light N$_2$ ASE for adaptive optics might well be realized as laser technology continues to improve.

\textcolor{black}{Other factors, in addition to LGS return flux, also affect AO performance. With large-aperture telescopes, LGS appear elongated for subapertures of the pupil that are offset from the axis of the laser launch telescope. This increases noise and reduces the wavefront-sensing accuracy. In this regard, an atomic species that has a high centroid altitude and small vertical extent would be favored. In practice, the altitude is limited by the rapidly decreasing atmospheric density and there is not a large difference between the atomic species listed in Table \ref{tab:Atomic Parameters}. If an LGS can be created by a high-power pulsed laser, an important consideration will be to limit the extent of the emitting region along the line of sight.}

\section{Summary and conclusions}

We have reviewed transitions of atoms and ions in the upper atmosphere of interest for AO.  Besides Na, which is used extensively by current systems, Fe, Mg, Si, Ca, K all have potential uses. Iron has the highest abundance and the largest number of transitions. However, its return flux is less \textcolor{black}{than} that of sodium due to lower cross sections.


Our results confirm the general presumption of the LGS community that Fe is not as well-suited as Na for PLGS. However, we find that one-step excitation at 330 nm may be more suitable than two-step excitation at $569 + 589$ nm. Detailed simulations, using Bloch equations and including the effects of Zeeman splitting, would be a useful next step to evaluate the potential of these excitation schemes.   

%
%

Mg$^+$ and Si$^+$ have high abundance and several very-strong visible-light transitions. However they require vacuum ultraviolet photons for excitation that appear to be implausible even with two-photon excitation. 

Amplified spontaneous emission using metallic species appears to be unfeasible because of the small optical depths. A high abundance is required for ASE, so nitrogen and oxygen are likely the only atoms that could be used. 

\begin{appendices}

\setcounter{equation}{0}
\renewcommand{\theequation}{A\arabic{equation}}

\section{Atomic collision rates in the mesosphere and lower thermosphere}

We wish to estimate the collision rates for various metallic atoms and ions in the mesosphere and lower thermosphere (MLT) region. The atmospheric composition in the MLT is essentially the same as in the troposphere, consisting almost entirely of three species, N$_2$, O$_2$ and Ar. Their fractions, by volume, are approximately 78\%, 21\% and 1\% respectively. The species of interest to us are Na, Fe, Ca, Si, Mg, K, N, O, H and their ions. 

The collision rate $R$, for a given atom, is given in terms of the collision cross-section Q and the number densities $n$ for the dominant atomic and molecular species. For atom $i$, interacting with field atoms or molecules $j$,
\be
  R_i = \sum_j n_j Q_{ij} v_{ij},
\ee
where $v_{ij}$ is the relative velocity. For a Maxwell-Boltzmann distribution at temperature $T$, the RMS relative velocity of an atom or molecule of mass $m_i$ and one of mass $m_j$ is,
\be
  v_{ij} = \sqrt{v_i^2+v_j^2} = \sqrt{\frac{8kT}{\pi \mu_{ij}}},
\ee
where $\mu_{ij} = m_im_j/(m_i+m_j)$ is the reduced mass. The low-energy collision cross section between two atoms is well-approximated by the hard-sphere model, 
\be
  Q_{ij} = \pi (r_i+r_j)^2,
\ee
where $r_i$ and $r_j$ are the atomic radii. For a diatomic molecule $j$, the situation is more complicated. Modelling the molecule by two spheres of radius $r_j$ in contact, the appropriate geometrical cross section is the projected area $A$ of two spheres of radius $r = r_i+r_j$, having centres separated by a distance $d$. The separation depends on the angle $\theta$ between the axis of the molecule and the velocity vector, $d = 2r_j\sin\theta$.  To find the mean cross-section we 
average over this angle, assuming an isotropic velocity distribution. From plane geometry, 
\be
  A = 2\pi r^2 - 2r^2\arccos\left(\frac{d}{2r}\right)+ \frac{d}{2}\sqrt{4r^2-d^2}.
\ee
The frequency with which a particular angle $\theta$ appears is proportional to $\sin\theta$, so
\begin{align}
  Q_{ij} & = \frac{4}{\pi}\int_0^{\pi/2} A \sin^2\theta d\theta, \nonumber \\
  & = 2\pi r^2 + \frac{r^4+2r^2r_j^2-3r_j^4}{\pi r_j^2}\text{arcsinh}\left(\frac{r_j}{\sqrt{r^2-r_j^2}}\right) \nonumber\\
  & \quad - \frac{r^3-3rr_j^2}{\pi r_j}
  +\frac{32r_jr}{3\pi \sqrt{2\pi}}{_3F_2}\left[2,\frac{1}{2},\frac{1}{2};\frac{5}{2},\frac{3}{2};\left(\frac{r_j}{r}\right)^2\right],
\end{align}
where $_3F_2$ is a generalized hypergeometric function.

Atomic radii are taken from \cite{Cox1999}. For the atmospheric density and temperature the  MSIS-E-90 atmospheric model was used \cite{Hedin1991}. \textcolor{black}{These were evaluated, for a latitude of $30^\circ$ ond longitude of $0^\circ$, at midnight on the first day of each month, and the average over a full year was computed.} The resulting mean cross sections for the atoms and ions of interest, colliding with N$_2$, O$_2$ and Ar, are listed in Table A1. The computed collision rates are listed in Tables A2 and A3. For comparison, Holzl\"ohner et al. \cite{Holzlohner2010} estimated a Na -- O$_2$ collision rate of 1/(35 $\mu$s) at an altitude of 92 km, \textcolor{black}{which is within a factor of two of our estimate}. 

\vspace{12pt}

\tablefirsthead{\multicolumn{6}{L{6.8cm}}{\bfseries Table A1. Mean collision cross sections.  }  \\[6pt] } 
\tablehead{\\[-7pt] \multicolumn{6}{l}{{\bfseries Table A1 continued}} \\  \hline} 
\tabletail{\\[-14pt]\midrule}
\tablelasttail{}
\begin{center}
\begin{supertabular}{lccccc}
\hline \\[-9pt]
Atom & mass & radius & \multispan{3}{\hfill cross section \hfill } \\
 & (amu) & (nm) & \multispan{3}{\hfill (nm$^2$)\hfill } \\
&&& N$_2$ & O$_2$ & Ar \\
\hline
Na & 22.990 & 0.191 & 0.565 & 0.550  & 0.277 \\
Fe & 55.845 & 0.127 & 0.315 & 0.303 & 0.171 \\
Mg & 24.305 & 0.162 & 0.443 & 0.429 & 0.226 \\
Si & 28.085 & 0.109 & 0.257 & 0.247 & 0.145 \\
Ca & 40.078 & 0.197 & 0.593 & 0.577 & 0.288 \\
K & 39.098 & 0.237 & 0.790 & 0.771 & 0.370 \\
N & 14.007 & 0.070 & 0.152 & 0.144 & 0.097 \\
O & 15.999 & 0.066 & 0.143 & 0.135 & 0.093 \\
Ar & 39.948 & 0.106 & 0.248 & 0.238 & 0.141 \\
H & 1.008 & 0.070 & 0.152 & 0.144 & 0.097 \\
\hline
\end{supertabular}
\end{center}

\tablefirsthead{\multicolumn{7}{L{7.5cm}}{\bfseries Table A2. Collision rates, in units of $10^2$ s$^{-1}$, for metallic species. }  \\[6pt] }
\tablehead{\\[-7pt] \multicolumn{7}{l}{{\bfseries Table A2 continued}} \\  \hline}
\tabletail{\\[-14pt]\midrule}
\tablelasttail{}
\begin{center}
\begin{supertabular}{Scccccc}
\hline 
{altitude} & Ca & Fe & K & Mg & Na & Si \\
{(km)} \\
\hline
 75 & 2449  & 1219  & 3283  & 2062  & 2675  & 1150  \\
 80 & 1108  & 551.7 & 1486  & 933.3 & 1211  & 520.7 \\
 85 & 484.1 & 241.0 & 649.1 & 407.7 & 528.8 & 227.5 \\
 90 & 203.7 & 101.4 & 273.1 & 171.5 & 222.5 & 95.70 \\
 95 & 82.66 & 41.15 & 110.8 & 69.60 & 90.28 & 38.84 \\
100 & 32.65 & 16.26 & 43.77 & 27.49 & 35.65 & 15.34 \\
105 & 13.08 & 6.515 & 17.54 & 11.01 & 14.28 & 6.147 \\
110 & 5.662 & 2.821 & 7.590 & 4.766 & 6.181 & 2.661 \\
115 & 2.740 & 1.365 & 3.673 & 2.306 & 2.991 & 1.288 \\
120 & 1.489 & 0.742 & 1.996 & 1.253 & 1.625 & 0.700 \\
\hline
\end{supertabular}
\end{center}

\tablefirsthead{\multicolumn{5}{L{5.8cm}}{\bfseries Table A3. Collision rates, in units of $10^2$ s$^{-1}$, for non-metallic species. }  \\[6pt] } 
\tablehead{\\[-7pt] \multicolumn{5}{l}{{\bfseries Table A3 continued}} \\  \hline} 
\tabletail{\\[-14pt]\midrule}
\tablelasttail{}
\begin{center}
\begin{supertabular}{Scccc}
\hline 
{altitude} & Ar & H & N & O \\
{(km)} \\
\hline
 75 & 1023  & 2592  & 833.3 & 748.7 \\
 80 & 463.0 & 1173  & 377.1 & 338.8 \\
 85 & 202.3 & 512.3 & 164.7 & 148.0 \\
 90 & 85.10 & 215.5 & 69.30 & 62.27 \\
 95 & 34.54 & 87.46 & 28.13 & 25.27 \\
100 & 13.64 & 34.54 & 11.11 & 9.982 \\
105 & 5.468 & 13.83 & 4.451 & 4.000 \\
110 & 2.367 & 5.984 & 1.926 & 1.731 \\
115 & 1.146 & 2.895 & 0.932 & 0.838 \\
120 & 0.623 & 1.573 & 0.507 & 0.455 \\
\hline
\end{supertabular}
\end{center}

\setcounter{equation}{0}
\renewcommand{\theequation}{B\arabic{equation}}

\section{Transition rates in a multi-level atom}

In a multi-level atom, with states $j$, $j = 0, 1,\cdots,n-1$, the fraction of atoms $x_j$  for which the electron is in state $j$ is determined by $n$ rate equations
\be
  \dot{x}_j = \Gamma_{jk} x_k.
\ee
Here $\Gamma_{jk}$ specifies the net transition rate between states $j$ and $k$. These equations are not linearly independent because the occupations fractions are related by the constraint
\be
  \sum x_k = 1.
\ee
This condition can be used to eliminate one of the variables, $x_0$ say, which results in a set of $n-1$ linearly-independent inhomogeneous equations
\be
  \dot{x}_j = M_{jk} x_k + b_j, \quad j = 1, 2, \cdots, n-1. \label{eq:rate}
\ee
The quantities $M_{jk}$ and $b_j$ depend on transition rates and the radiation energy densities at the transition wavelengths. If the radiation energy densities can be regarded as independent of time, these quantities are constant and the equations are readily solved using standard techniques. The general solution is  
\be
  \bs{x} =  \sum c_k \bs{a}_k e^{\lambda_k t} - M^{-1} \bs{b},
\ee
where $\lambda_k$ and $c_k$ are the eigenvalues and eigenvectors of $M$. 

The steady-state solution is found by equating the left hand side of Eqn (\ref{eq:rate}) to zero,
\be
  \bs{x} =  - M^{-1} \bs{b},
\ee

\end{appendices}

\noindent{\large\sffamily\bfseries Funding.}
Yunnan Provincial Department of Education; Natural Sciences and Engineering Research Council of Canada (RGPIN-2019-04369); Chinese Academy of Sciences, CAS President’s International Fellowship Initiative (2017VMA0013). \\[6pt]

\noindent{\large\sffamily\bfseries Acknowledgement.}
We thank Profs. D. Budker and R. Holzl\"ohner for comments on an earlier version of the manuscript. PH thanks the National Astronomical Observatories, Chinese Academy of Sciences for hospitality during a sabbatical visit. \\[6pt]

\noindent{\large\sffamily\bfseries Disclosures.}
The authors declare no conflict of interests. \\[6pt]

\noindent{\large\sffamily\bfseries Data availability.} 
Data underlying the results presented in this paper are available in Refs. \cite{Hedin1991} and \cite{Kramida2020}. Selected data for strong transitions in the atomic species discussed here are available from the authors upon request.


\noindent 

\begin{thebibliography}{1}

\bibitem{Beckers1993}
J. M. Beckers, 
 Ann. Rev. Astron. Astrophys., \textbf{31}, 13
 (1993).

\bibitem{Foy1985}
 R. Foy and A. Labeyrie, 
 Astron. Astrophys. \textbf{152}, L29
 (1985).
 
\bibitem{Happer1994}
W. Happer, G. J. MacDonald, C. E. Max and F. J. Dyson, 
 J. Opt. Soc. Am. A, \textbf{11}, 263
 (1994).
 
\bibitem{Telle1996}
J. M. Telle,
Opt. Soc. Am. Tech. Dig. Ser. \textbf{13}, 100
(1996).
 
\bibitem{Fan2016}
T. W. Fan, T. H. Zhou, and Y. Feng,
Sci. Rep. \textbf{6},19859 (2016).

\bibitem{Holzlohner2010}a
R. Holzl\"ohner, S. M. Rochester, D. B. Calia, D. Budker, J. M. Higbie, and W. Hackenberg,
Astron. Astrophys. \textbf{510}, A20 (2010).
   
\bibitem{Milonni1999}
P. W. Milonni, H. Fearn, J. M. Tell and R. Q. Fugate,
J. Opt. Soc. Am. A \textbf{16}, 11 (1999).
 
\bibitem{Rampy2015}
R. Rampy, D. Gavel, S. M. Rochester and R. Holzl\"ohner, 
J. Opt. Soc. Am. B \textbf{32}, 2425
(2015).

\bibitem{Bustos2020}
F. P. Bustos, R. Holzl\"ohner, S. Rochester, D. B. Calia, J. Hellemeier and D. Budker,  
J. Opt. Soc. Am. B \textbf{37}, 1208
(2020).

\bibitem{Hickson2020}
P. Hickson, J. Hellemeier and R. Yang, 
Opt. Lett., \textcolor{black}{\textbf{46}, 1792}
(2021).  
 
\bibitem{Bustos2018}
F. P. Bustos, A. Akulshin, R. Holzl\"ohner, et al., 
Proc. SPIE \textbf{10703}, 107030R-1 (2018).
 
\bibitem{Rigaut1992}
F. Rigaut and E. Gendron,
 Astron. Astrophys. \textbf{261}, 677
 (1992).
 
\bibitem{Esposito1998}
\textcolor{black}{
S. Esposito,
 Proc. SPIE. \textbf{3353}, 468
 (1998).
 }

\bibitem{Ragazzoni2000}
\textcolor{black}{
R. Ragazzoni,
 ``Laser Guide Star Adaptive Optics for Astronomy,'' Ed. N. Ageorges and C. Dainty, NATO ASI Series C, \textbf{551}, 125
 (2000).
}

\bibitem{Foy1995}
R. Foy, A. Migus, F. Biraben, G. Grynberg, P. R. McCullough, and M. Tallon,
Astron. Astrophys. Suppl. \textbf{111}, 569
(1995).

\bibitem{Pique2006}
J. P. Pique, I. C. Moldovan, and V. Fesquet,
J. Opt. Soc. Am. A \textbf{23}, 2817
(2006).

\bibitem{Pfrommer2010}
\textcolor{black}{
Pfrommer, T. and Hickson, P., 
JOSA A, \textbf{27}, A97
(2010)
}

\bibitem{Davis2006}
\textcolor{black}{
D. S. Davis, P. Hickson, G. Herriot, and C-Y She, 
Optics Let., \textbf{31}, 3369
(2006).
}

\bibitem{She2000}
C. Y. She, S. Chen, Z. Hu, J. Sherman, J. D. Vance, V. Vasoli, M. A. White, J. Yu, and D. A. Krueger,
Geophys. Res. Let. \textbf{27}, 3289
(2000).
 
\bibitem{Gardner2004}
C. S. Gardner,
App. Opt. \textbf{43}, 4941
(2004).

\bibitem{Hedin1991}
A. E. Hedin,
J. Geophys. Res. \textbf{96}, 1159 (1991), available online at \textcolor{black}{https://ccmc.gsfc.nasa.gov/modelweb}
 
\bibitem{Whalley2011}
C. L. Whalley, J. C. Gomez-Martin, T. G. Wright, et al.,
Phys. Chem. Chem. Phys., \textbf{13}, 6352
(2011).

\bibitem{Plane2010}
\textcolor{black}{J. M. C. Plane, 
Air Force Research Laboratory Report AFRL-AFOSR-UK-TR-2010-0005, 22 pp,
(2010).}
 
\bibitem{Plane2016}
 J. M. C. Plane, J. C. Gomez-Martin, W. Feng, and D. Janches,
J. Geophys. Res. Atmos. \textbf{121}, 3718
(2016).

\bibitem{Aikin2000}
 A. C. Aikin and H. J. P. Smith,
Phys. Chem. Earth (C) \textbf{25}, 203
(2000).
  
\bibitem{Mlynczak2013}
M. G. Mlynczak, L. A. Hunt, J. C. Mast, et al., 
J. Geophys. Res. Atmos. \textbf{118}, 5724
(2013).  
 
\bibitem{Mlynczak2014}
M. G. Mlynczak, L. A. Hunt, B. T. Marshall, et al.,
J. Geophys. Res. Atmos. \textbf{119},  3516
(2014). 

\bibitem{Kramida2020}
A. Kramida, Y. Ralchenko, J. Reader, J. and NIST ASD Team (2020). NIST Atomic Spectra Database (ver. 5.8), [Online]. Available: https://physics.nist.gov/asd [2020, November 17]. National Institute of Standards and Technology, Gaithersburg, MD. DOI: https://doi.org/10.18434/T4W30F

\bibitem{Marsh2013}
D. R. Marsh, D. Janches, W. H Feng, and J. M. C. Plane,
J. Geophys. Res. Atmos. \textbf{118},11442
(2013).

\bibitem{Pique2003}
\textcolor{black}{
J-P Pique and S. Farinotti, 
JOSA B \textbf{20}, 2093
(2003).
}

\bibitem{Biegert2003}
J. Biegert and J. C. Diels,
Phys. Rev. A \textbf{67}, 043403 (2003).
 
\bibitem{Xu2017}
C. Xu, C. Guo, H-B. Yu, et al.,
Appl. Phys. B \textbf{123}, 94 (2017). 
 
\bibitem{Feng2013}
W. H. Feng, D. R. Marsh, M. P. Chipperfield, et al.,
J. Geophys. Res. Atmos. \textbf{118}, 9456
(2013). 
 
\bibitem{Kelleher2008a}
D. E. Kelleher and L. I. Podobedova
J. Phys. Chem. Ref. Data \textbf{37},  267
(2008).
 
\bibitem{Kelleher2008b}
D. E. Kelleher and L. I. Podobedova,
J. Phys. Chem. Ref. Data \textbf{37}, 1285
(2008).
 
\bibitem{Plane2011}
 J. M. C. Plane,
Nature Chem. \textbf{3}, 900  (2011).
 
\bibitem{Kopp1997}
E. Kopp,
J. Geophys. Res. \textbf{102}, 9667
(1997).
 
\bibitem{Maurer2004}
C. Maurer, C. Becher, C. Russo, J. Eshner, and R. Blatt,
New J. Phys. \textbf{6}, 94 (2004).
  
\bibitem{Conti2001}
S. Conti, P. I. Porshnev, A. Fridman, et al.,
Exp. Therm. Fluid Sci. \textbf{14}, 79
(2001).
 
\bibitem{Cochran2001}
 A. L. Cochran and W. D. Cochran,
Icarus \textbf{154}, 381
(2001).
  
\bibitem{Grossmann2006}
K. U. Grossmann, O. Gusev, and P. Knieling,
J. Atmos. Solar-Terr. Phys. \textbf{68}, 1764
(2006). 
 
\bibitem{Rairoux2000}
 P. Rairoux, H. Schillinger, S. Niedermeier, et al.,
Appl. Phys. B \textbf{71}, 573
(2000).

\bibitem{Cox1999}
A. N. Cox, ``Allen's Astrophysical Quantities,'' Springer, NY (1999).

\end{thebibliography}
\end{document}